\documentclass[aps,pra,twocolumn,notitlepage,nofootinbib,superscriptaddress]{revtex4-1}
\usepackage{xcolor}
\usepackage{graphicx}
\usepackage{rahulbhadani_arxiv}
\newcommand{\CEA}{C_{\text{EA}}\xspace}
\newcommand{\OH}{\text{OH}\xspace}
\newcommand{\OPC}{\text{OPC}\xspace}
\newcommand{\signal}{\ahat_s\xspace}
\newcommand{\idler}{\ahat_i\xspace}
\newcommand{\Nth}{N_{\text{th}}}
\graphicspath{ {figures/} }

\usepackage{physics}
\definecolor{blue}{RGB}{0,0,255}

\def\BibTeX{{\rm B\kern-.05em{\sc i\kern-.025em b}\kern-.08em
 T\kern-.1667em\lower.7ex\hbox{E}\kern-.125emX}}

\begin{document}
\title{Optimized Receiver Design for Entanglement-Assisted Communication using BPSK}

\author{Rahul Bhadani}
\affiliation{Department of Electrical and Computer Engineering, The University of Alabama in Huntsville, USA}
\email{rahul.bhadani@uah.edu, rahulbhadani@email.arizona.edu}
\author{Ivan B. Djordjevic}
\affiliation{Department of Electrical \& Computer Engineering, The University of Arizona, Tucson, USA\\College of Optical Sciences, The University of Arizona, Tucson, USA}
\email{ivan@email.arizona.edu}

\begin{abstract}
The use of pre-shared entanglement in entanglement-assisted communication offers a superior alternative to classical communication, especially in the photon-starved regime and highly noisy environments. In this paper, we analyze the performance of several low-complexity receivers that use optical parametric amplifiers. The simulations demonstrate that receivers employing an entanglement-assisted scheme with phase-shift-keying modulation can outperform classical capacities. We present a 2x2 optical hybrid receiver for entanglement-assisted communication and show that it has a roughly 10\% lower error probability compared to previously proposed optical parametric amplifier-based receivers for more than 10 modes. However, the capacity of the optical parametric amplifier-based receiver exceeds the Holevo capacity and the capacities of the optical phase conjugate receiver and 2x2 optical hybrid receiver in the case of a single mode. The numerical findings indicate that surpassing the Holevo and Homodyne capacities does not require a large number of signal-idler modes. Furthermore, we find that using unequal priors for BPSK provides roughly three times the information rate advantage over equal priors.
\end{abstract}

\maketitle

%%%%%%%%%%%%%%%%%%%%%%%%%%  body  %%%%%%%%%%%%%%%%%%%%%%%%%%
\section{Introduction}
\label{sec:intro}
Quantum Information Processing (QIP) has seen tremendous progress in recent decades, with multiple research directions exploring quantum sensing, covert communication, quantum cryptography, and more. A quantum channel is used to transfer quantum information from one party (known as Alice) to another party (known as Bob). In the case of a perfect channel, the quantum information is transferred intact, but if the channel is noisy, the quantum information undergoes some changes. Quantum channels can also be used to transmit classical information. Additionally, if the channel is noisy within certain limitations, the quantum channel can be used to share entanglement between Alice and Bob. The use of pre-shared entanglement can enhance classical capacity and protect against an adversary, commonly referred to as Eve~\cite{holevo2001evaluating, holevo2002entanglement, bennett2002entanglement, shi2020practical, zhuang2021quantum}. Recent experiments have shown that even in entanglement-breaking scenarios, the rate of entanglement-assisted (EA) communication can be much higher than communication without entanglement~\cite{zhang2015entanglement, hao2021entanglement}. The ratio $\frac{\CEA}{C}$, where $\CEA$ is the entanglement-assisted capacity and $C$ is the Holevo-Schumacher-Westmoreland (HSW) capacity in the classical regime, diverges logarithmically with the inverse of the signal power over a lossy and noisy bosonic channel~\cite{holevo2002entanglement}.

Recent efforts have been made to design receivers for EA communication, where authors have utilized the Gaussian approximation of the cumulative distribution function to calculate the Bit Error Rate (BER)~\cite{guha2009receiver, shi2020practical, zhuang2021quantum, djordjevic2021quantum}. The previously proposed receiver design is limited to a demonstration using Binary Phase-Shift Keying (BPSK) with repetition coding over more than $10^6$ bosonic modes that occupy the entire C-band and a portion of the L-band.

In this work, we analyze the receiver design for entanglement-assisted (EA) communication using Optical Parametric Amplifiers (OPAs) introduced in~\cite{djordjevic2021quantum} and expand upon previous results to determine the optimality of the receiver design. We show that EA communication does not need to occupy the entire C-band. Additionally, we analyze a 2x2 optical hybrid-based receiver for EA communication that is suitable for implementation in integrated optics and quantum nanophotonics. In our scheme, optical phase conjugation is performed on the transmitter side when signal photons are brighter, rather than on the receiver side where the signal photons are buried in noise and highly attenuated. A comparison of phase conjugation on the transmitted side versus the receiver side can be found in~\cite{djordjevic2021entanglement}.

We further propose an optimized hypothesis testing scheme and demonstrate numerically that the optimized receiver design provides a superior communication capacity compared to capacity without entanglement assistance. When using the \newchange{BPSK} modulation format to represent digital information, we find that non-equal priors perform at least three times better in terms of information rate compared to an equal prior encoding scheme. The development presented in this work is an extension of~\cite{bhadani2021optimal}.

The rest of the paper is organized as follows. In Section \ref{sec:review_entanglement}, we provide a brief review of entanglement-assistance with mathematical formalism necessary for the rest of the paper. In Section~\ref{sec:receiver_design}, we present an overview of the receiver design schemes for entanglement-assisted communication, including the optical parametric amplifier-based receiver design with threshold detection, the optical phase conjugation receiver, and the 2x2 optical hybrid-based joint receiver proposed in previous work~\cite{djordjevic2021quantum}. These receiver designs are then evaluated in Section~\ref{sec:evaluation}.

\subsection*{Notations Used in the Paper}
$\ket{\cdot}$ is used for ket-notation in quantum mechanics, equivalent to a vector notation in linear algebra. The Hermitian conjugate of the vector, $\bra{\cdot}$, is referred to as bra-notation. The scalar product of two vectors $\ket{\psi_1}$ and $\ket{\psi_2}$ is denoted by $\bra{\psi_1}\ket{\psi_2}$. Additionally, the ket-notation $\ket{\alpha}$ represents a coherent state of amplitude $\alpha$. The imaginary unit or a complex number $\sqrt{-1}$ is represented by $j$. Random variables $X$ and $Y$ denote the input and detected states, respectively. The measurement operator is represented by $\Pi$. Shannon's entropy is denoted by $H(\cdot)$ and mutual information is represented by $I(\cdot,\cdot)$. Probabilities are written as $p$, while conditional probabilities are represented as $p_{Y|X}$ and conditioned on $Y$ given $X$. The binomial coefficient is represented by $\binom{M}{N}$. The tensor product is represented by $\otimes$ and the cumulative distribution function of a statistical distribution is represented by $\mathcal{F}$.

\section{Entanglement Assisted Classical Communication Concept}
\label{sec:review_entanglement}
Quantum entanglement is a phenomenon where two particles are strongly correlated, such that the state of one particle immediately provides information about the state of the other particle, no matter how far apart they are. These particles, such as photons or electrons, are individual systems, but they remain connected even when separated by vast distances, forming a composite system~\cite{plenio1998teleportation,shao2010quantum,eisert2006entanglement}. As an example, given two basis vectors $\{ \ket{0}_A,\ket{1}_A \}$ in Hilbert space $\Hcal_A$ and $\{ \ket{0}_B, \ket{1}_B \}$ in Hilbert space $\Hcal_B$, then the following is an entangled state:
\spliteq{
\label{eq:entangled_state}
\cfrac{1}{\sqrt{2}} ( \ket{0}_A \otimes \ket{1}_B - \ket{1}_A \otimes \ket{0}_B )
}
When a composite system is in the state \eqref{eq:entangled_state}, it is impossible to attribute either system $A$ or $B$ a definite pure state. Although the von Neumann entropy of the whole state is zero, the entropy of the subsystem is greater than zero, indicating the systems are entangled.

Compared to classical communication, entanglement enhances communication by increasing the number of messages that can be sent perfectly over the channels, resulting in higher one-shot zero-error capacity and increased security~\cite{miller2011entanglement}. \newchange{However,}~\cite{miller2011entanglement}\newchange{doesn't explain what kind of measurement device and receiver scheme the experimentalists used.}  Theoretical proofs and discussions of entanglement-assisted communication can be found in~\cite{bennett1999entanglement, giovannetti2003entanglement, cubitt2010improving}. A laboratory experiment demonstrating the superiority of entanglement-assisted communication was recently conducted in~\cite{hao2021entanglement}.

\begin{figure}[htpb]
\centering
\includegraphics[width=0.75\linewidth, trim={0cm 0cm 0cm 0cm},clip]{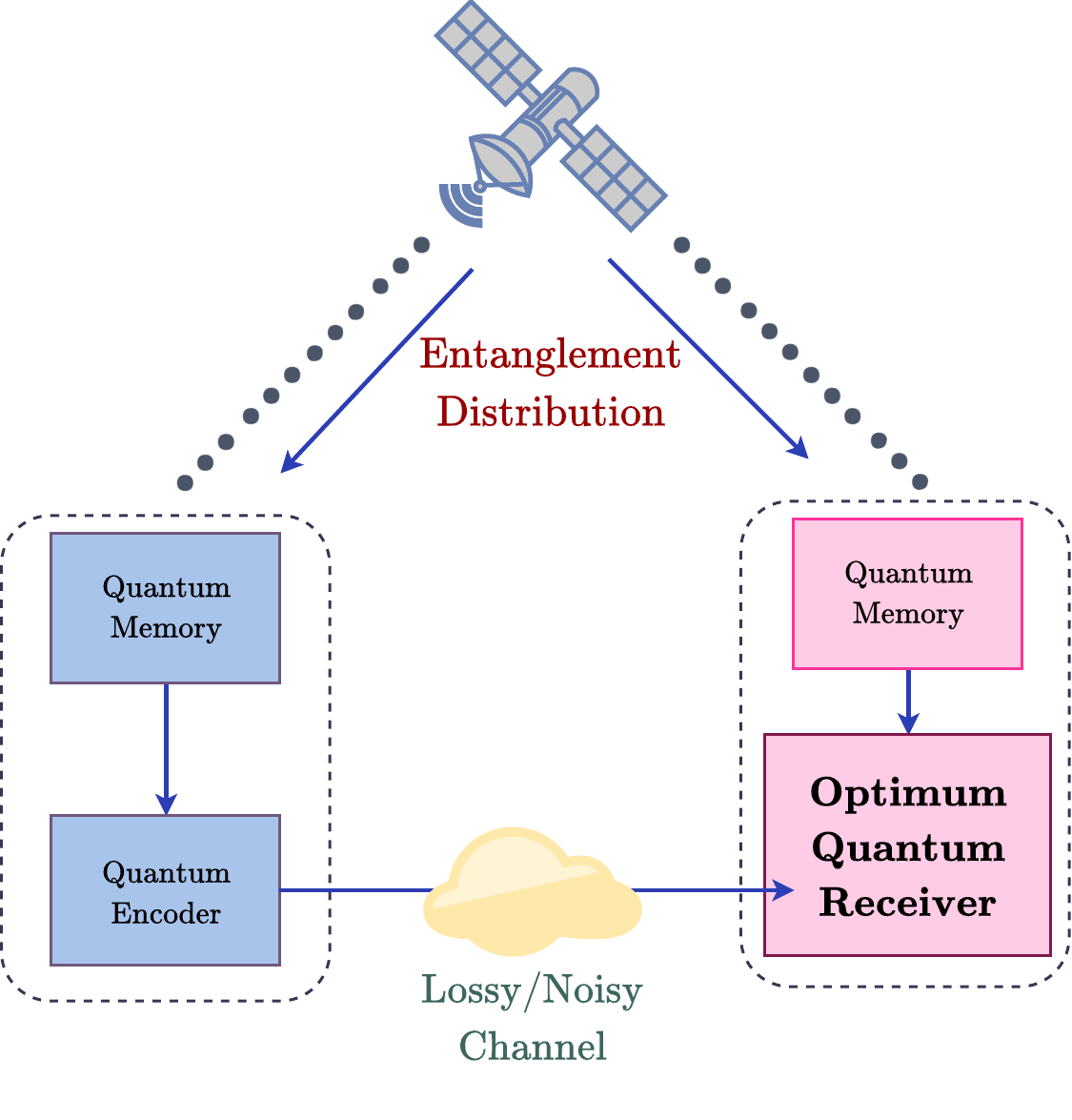}
\caption{An illustration showing the concept of EA communication. EA-assisted communication is enabled by either fiber-optic-based distribution of entanglement or entanglement distribution through satellites. The information is transmitted through a lossy and noisy Bosonic channel. }
\label{fig:EA_illustration}
\end{figure}

In entanglement-assisted classical communication, entangled states can be distributed through either optical fibers or satellites and stored in quantum memories. The classical data is transmitted by Alice using the signal photon of the entangled pair, which is affected by noise and loss in the quantum channel. On the receiver side, Bob uses the idler photon of the entangled pair to determine what was transmitted by employing an optimal quantum receiver. The overall design is illustrated in Figure \ref{fig:EA_illustration}. Error correction can be applied to the quantum states to restore the transmitted information and mitigate the effects of decoherence.

\section{Receiver Design for EA Communication}
\label{sec:receiver_design}

In entanglement-assisted communication, two-mode Gaussian states are generated through spontaneous parametric down-conversion (SPDC) of entangled-photon pairs~\cite{lopez2012implementation, CHRIST2013351}. 
% A brief discussion of SPDC can be found in Appendix~\ref{sec:spdc}.
The SPDC source is a broadband source \newchange{with a number of modes} $M = T_m W$, where $W$ is the phase-matching bandwidth and $T_m$ is the measurement interval and generates $M$ independent pairs of signal-idler photons in space and time denoted by their annihilation operators ${ \signal^{(m)}, \idler^{(m)}}$, with $m \in [1, M]$. These pairs are prepared in identical entangled two-mode squeezed vacuum (TMSV) states, which can be represented in a Fock state basis as in Equation~\eqref{eq:tmsv_fock}
\spliteq{
\label{eq:tmsv_fock}
\ket{\psi}_{si} = \sum_{n=0}^{\infty}\sqrt{\cfrac{N_s^n}{(N_s +1)^{n+1} }}\ket{n}_s\ket{n}_i
}
where $N_s$ is the mean photon number in the signal mode. The mean photon number for the idler mode is $N_i = N_s$~\cite{mauerer2009colors, guha2009receiver}. TMSV belongs to a class of Gaussian states, where an $M$-mode Gaussian state $\hat{\rho}$ consisting of modes ${\ahat^{(m)}, m \in [1, M]}$ is characterized by the mean and variance of their respective quadrature field operators such that $\ahat^{(m)} = \phat^{(m)} + j \qhat^{(m)}$. The covariance matrix for a TMSV state is given by
\spliteq{
\label{eq:COVTMSV}
\Lambda_{\text{TMSV}} & = \m{ 2N_s + 1 & 0  &  C & 0 \\ 0 & 2N_s + 1 & 0 & -C\\ C & 0 & 2N_s + 1 & 0 \\ 0 & -C & 0 & 2N_s + 1 }\\
& = \m{(2N_s + 1)\Ibf & C \Zbf \\ C \Zbf  & (2N_s + 1)\Ibf}
}
where $C = 2\sqrt{N_s (N_s +1)}$, $\Ibf$ and $\Zbf$ are $2\times 2$ Pauli matrices. Other two Pauli matrices are $\Xbf$ and $\Ybf$\footnote{\url{https://qiskit.org/documentation/stubs/qiskit.quantum_info.Pauli.html}}~\cite{wilde2013quantum}.

\newchange{If we consider the Phase-Shift Keying (PSK) modulation scheme for communication, then mathematically, we can use the unitary operator $\Uhat_\theta = e^{j\theta \ahat^\dagger \ahat}$ to denote the rotation of the base annihilation operator $\ahat$. In transmitting information using entangled photons generated from SPDC, the signal photon of the signal-idler pair is used while the idler is pre-shared before transmission occurs. 
In order to transmit information, Alice modulates the signal $\ahat_{s'}$ using a phase modulator to apply a rotation of $\theta$. The signal then passes through a thermal, lossy Bosonic quantum channel. The received photon mode (after passing through the communication channel) at Bob's end is denoted by $\ahat_R = \ahat_{R'}e^{j\theta}$ where $\ahat_{R'}$ is the base photon mode at the receiving end. Bob uses the undisturbed idler part of the pre-shared entangled photon pair and an optimal quantum detector to perform hypothesis testing and determine which symbol was transmitted.} For simplicity, we will drop the mode notation from the annihilation operator. Under the phase-encoding scheme, the covariance matrix of the return-idler pair ${\ahat_R, \ahat_I }$ is given by
\spliteq{
\label{eq:cov_signal_idler}
\m{(2N_R + 1)\Ibf & C_\eta  Re[e^{j\theta} (\Zbf - j\Xbf)] \\ C_\eta Re[e^{j\theta} (\Zbf - j\Xbf)]  & (2N_I + 1)\Ibf}
} 
where $N_R=\eta N_s+N_B$, $C_\eta = 2\sqrt{\eta N_s (N_s +1)}$, $\eta$ is the transmittivity of the Bosonic channel, and $N_B$ is the mean photon number of the thermal mode. \newchange{In the case of a pre-shared entangled state, the idler is assumed to be undisturbed as it has been shared through fiber optics or satellite and is stored in quantum memory. In such a case, attenuation experienced by the idler is negligible. Hence, at the receiver side, the idler mean photon number $N_I = N_i = N_s$. As the signal mode passes through a thermal lossy bosonic channel, the signal mode is altered and referred to as the return mode on the receiver side with mean photon number $N_R$.}

\subsection{OPA-based receiver with threshold detection}
\label{sec:OPA}

\begin{figure}[htpb]
\centering
\includegraphics[width=0.95\linewidth, trim={0cm 0cm 0cm 0cm},clip]{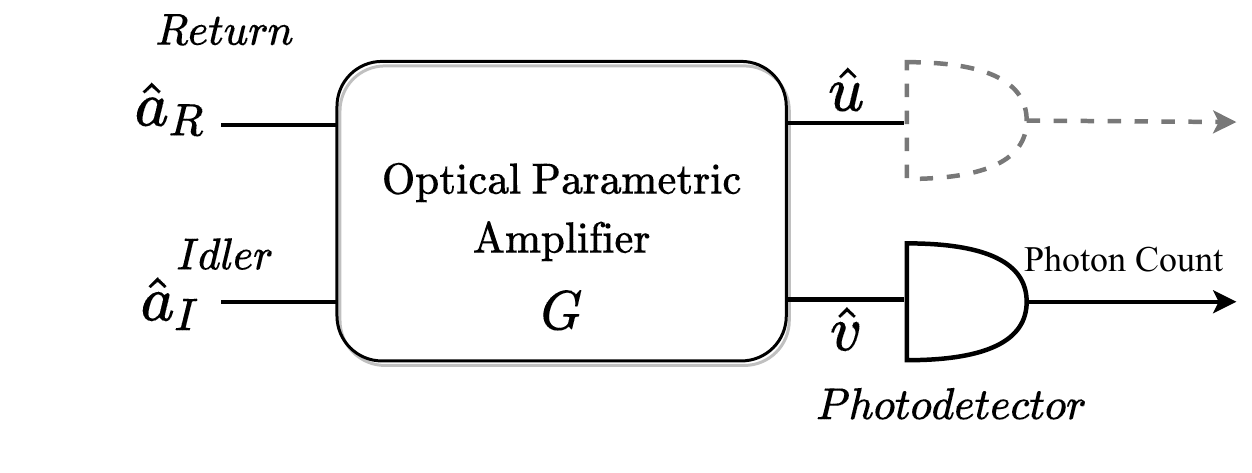}
\caption{Operating Principle of Optical Parametric Amplifier (OPA): At the receiver end, parametric amplification is applied to the return-idler pair with gain $G$. Error probability of discrimination is higher at $\uhat$, hence the photon detection is made $\vhat$.}
\label{fig:OPA2}
\end{figure}

A joint detection receiver for state discrimination of EA communication consists of an optical parametric amplifier (OPA). On the receiver side, an optical parametric amplifier (OPA) is used to combine the return-idler pair, as shown in Figure~\ref{fig:OPA2}. The return and idler modes are evolved as given by Heisenberg's picture~\cite{ou2017quantum}:
\spliteq{
\label{eq:OPA_Combine}
\uhat & = \sqrt{G}\ahat_R + \sqrt{G-1}\ahat_I^\dagger\\
\vhat & = \sqrt{G}\ahat_I + \sqrt{G-1}\ahat_R^\dagger
}
where $G$ is the gain of the OPA, such that $G = 1 + \epsilon$ and $\epsilon << 1$.
OPA receiver can be used to combine and amplify the return-idler pair using a strong local pump. This gives rise to Equation~\eqref{eq:OPA_Combine}. At the output ports, a photodetector is used for photon counting, and a threshold detection rule is applied to make state discrimination. We further assume an ideal OPA, where the gain $G$ is fixed.

The photodetector outputs are designated as $\uhat$ and $\vhat$ at two ports, referred to as the return and idler outputs, respectively. For each output, the mean photon number is given by the expectation $\expval{\uhat^\dagger \uhat}$ or $\expval{\vhat^\dagger \vhat}$, depending on whether threshold detection is made at the signal output port or idler output port. The photocurrent operators and their expectations are given by Equation~\eqref{eq:OPA_meanphotonOutput} and Equation~\eqref{eq:OPA_meanphotonOutput2}.
\spliteq{
\label{eq:OPA_meanphotonOutput}
\Nbar_1(\theta) & = \expval{\uhat^\dagger \uhat} = G(\eta N_s + N_B) \\& + (G-1)(1 + N_s)  + 2\cos\theta \sqrt{G(G-1)}\sqrt{\eta N_s (N_s + 1)}
}
\spliteq{
\label{eq:OPA_meanphotonOutput2}
\Nbar_2(\theta) & = \expval{\vhat^\dagger \vhat}  =  GN_s + (G-1)( 1 + \eta N_s + N_B) \\&  + 2\cos\theta \sqrt{G(G-1)}\sqrt{\eta N_s (N_s + 1)} 
}
The derivation is provided in Appendix~\ref{sec:opa_mean_derivation}. 

For practical communication, consider that information is encoded using repetition codewords that employ binary phase-shift keying (BPSK) modulation with phases $\theta \in {0, \pi}$. Decoding BPSK can be modeled as hypothesis testing: if hypothesis $H_0$ is true, then the BPSK symbol with $\theta = 0$ was transmitted, and if hypothesis $H_1$ is true, then the symbol with $\theta = \pi$ was transmitted. In this paper, we do not discuss optimal encoding, which is beyond the scope of this paper. However, BPSK is a suitable choice for weak signals, as it is power-efficient~\cite{lefevre1989power}.

To allow for efficient error correction, repeated PSK codewords consisting of $M$ signal-idler pairs are used in EA communication~\cite{shi2020practical}. In a joint-detection scheme, the receiver mixes all $M$ received modes and counts the total number of photons at the output ports. The joint detection state in this case becomes an $M$-fold tensor product $\rho^{\otimes M}$, with identical zero-mean thermal states, and the per-mode mean photon number is given by $\Nbar_1(\theta)$ or $\Nbar_2(\theta)$, depending on which output port of the OPA we use. An optimum joint measurement for state discrimination requires photon counting at an output port and thus deciding between two hypotheses using the total photon number $N$ over $M$ modes~\cite{guha2009receiver, tan2008quantum}. Under such a scenario, the probability mass function (pmf) is negative binomial with mean $M\Nbar(\theta)$ and standard deviation $\sigma(\theta) = \sqrt{M\Nbar(\theta)(\Nbar(\theta) + 1)}$, given by~\cite{guha2009receiver, tan2008quantum}:
\spliteq
{
\label{eq:OPA_P}
P_\text{OPA}(n|\theta; M, i) = \binom{n+M-1}{n}\bigg(\tfrac{\Nbar_i(\theta)}{1 + \Nbar_i(\theta)}\bigg)^n  \bigg(\tfrac{1}{1 + \Nbar_i(\theta)}\bigg)^M
}
where $i \in {1, 2}$, and $\binom{n+M-1}{n}$ is the binomial coefficient.

Equation \eqref{eq:OPA_P} can be approximated as a Gaussian distribution with mean $M\Nbar_i(\theta)$ and standard deviation $\sigma(\theta)=\sqrt{M\Nbar_i(\theta)(\Nbar_i(\theta)+1)}$ for sufficiently large $M$ (see Appendix~\ref{sec:opa_optimum_nth}). At the detector end, we use threshold detection and decide in favor of $H_0$ if the total number of photons detected is $N>\Nth(\theta)$, otherwise we choose $H_1$ for $N\leq\Nth(\theta)$, where $\Nth(\theta)$ is the threshold number of photons, which is a function of the phase $\theta$. A suitable value for the threshold number of photons is chosen according to the scheme described later in Section~\ref{sec:evaluation}.

\subsection{Optical Phase Conjugation Receiver with Threshold Detection}

\begin{figure}[htpb]
\centering
\includegraphics[width=1.0\linewidth, trim={2cm 0cm 0cm 0cm},clip]{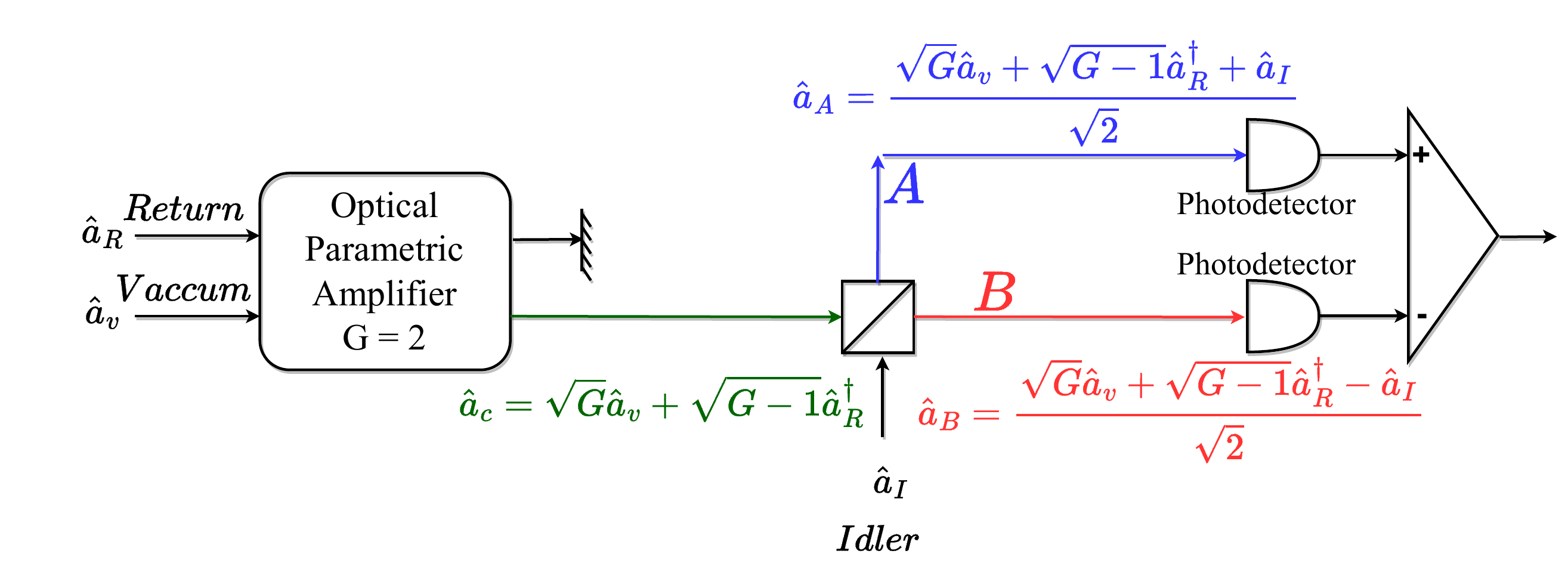}
\caption{Operating Principle of Optical Phase Conjugate (OPC) receiver: signal interacts with the vacuum followed by mixing with idler using 50-50 beamsplitter and a balanced detection is applied using photodetectors.}
\label{fig:OPC}
\end{figure}

OPA can also be used differently, where the return $\ahat_R$ mode interacts with the vacuum mode $\ahat_v$ to produce $\sqrt{G}\ahat_v + \sqrt{G-1}\ahat_R^\dagger$, which becomes $\ahat_c=\sqrt{2}\ahat_v + \ahat_R^\dagger$ for $G=2$. By mixing the idler with $\ahat_c$ using a 50-50 beamsplitter, we get two modes $\cfrac{1}{\sqrt{2}}(\ahat_c\pm\ahat_I)$. The outputs from the two arms are fed to a balanced detector, and their difference is measured as a photocurrent. We call this the Optical Phase Conjugate Receiver (OPC receiver). Consider the schematic of the OPC receiver shown in Figure~\ref{fig:OPC}.

For the case of BPSK, the mean photon operators of two output arms of beamsplitters are given by 
{
\small{
\spliteq
{
\label{eq:OPC_meanoperator}
\ahat_{A/B}^\dagger\ahat_{A/B} & = \cfrac{1}{2}\bigg[ (G-1)\ahat_R \ahat_R^\dagger \pm \sqrt{G-1}\ahat_R\ahat_I  \\ & \pm \sqrt{G-1}\ahat_I^\dagger\ahat_R^\dagger + \ahat_I^\dagger \ahat_I \bigg]\\
& \text{with $+$ sign for arm A and $-$ sign for arm B.}
}
}
}
In Equation~\eqref{eq:OPC_meanoperator}, $\ahat_v$ term doesn't appear as it denotes the vacuum mode.

We adopt a joint-detection scheme similar to the one adopted for the OPA receiver discussed in Section~\ref{sec:OPA}, containing $M$ modes for error correction. The difference in the mean photon number detected at the two photodetectors of the OPC is converted to a photocurrent with a photocurrent operator $\khat$ given by Equation~\eqref{eq:OPCMean}, \newchange{setting $G = 2$}.
\spliteq
{
\label{eq:OPCMean}
\khat & = \ahat_{A}^\dagger\ahat_{A} - \ahat_{B}^\dagger\ahat_{B} = \sqrt{G-1}\ahat_R\ahat_I + \sqrt{G-1}\ahat_I^\dagger\ahat_R^\dagger \\
N_{\text{OPC}}(\theta) & = \expval{\khat} = 2\cos \theta\sqrt{\eta Ns(Ns + 1)}\\
& \text{as,~} \ahat_R = \ahat_{R'}e^{j\theta},\\ & ~\expval{\ahat_{R'} \ahat_I} = \sqrt{\eta N_s (N_s + 1)},~\ahat_R \ahat_R^\dagger = \ahat_R^\dagger \ahat_R + I
}
The variance $\sigma^2_{\text{OPC}}$ is given by Equation~\eqref{eq:OPCVar}, \newchange{setting $G = 2$}.
\spliteq
{\label{eq:OPCVar}
\sigma^2_{\text{OPC}}(\theta) & = \expval{\khat^2} - \expval{\khat}^2 \\
& = N_s(\eta N_s  + N_B + 1 ) + (N_s + 1)(\eta N_s + N_B + 1) \\ & - 2 (\eta N_s (N_s + 1 ) ) \cos2\theta - 4 ( \eta N_s (N_s + 1) ) \cos^2\theta
}
At the detector end, the decision scheme uses threshold detection, similar to the OPA-based receiver design discussed in Section~\ref{sec:OPA}.

\subsection{2x2 Optical Hybrid-based Joint Receiver with Threshold Detection}
\label{eq:OH2x2}
\begin{figure}[htpb]
\centering
\includegraphics[width=0.75\linewidth, trim={0cm 0cm 0cm 0cm},clip]{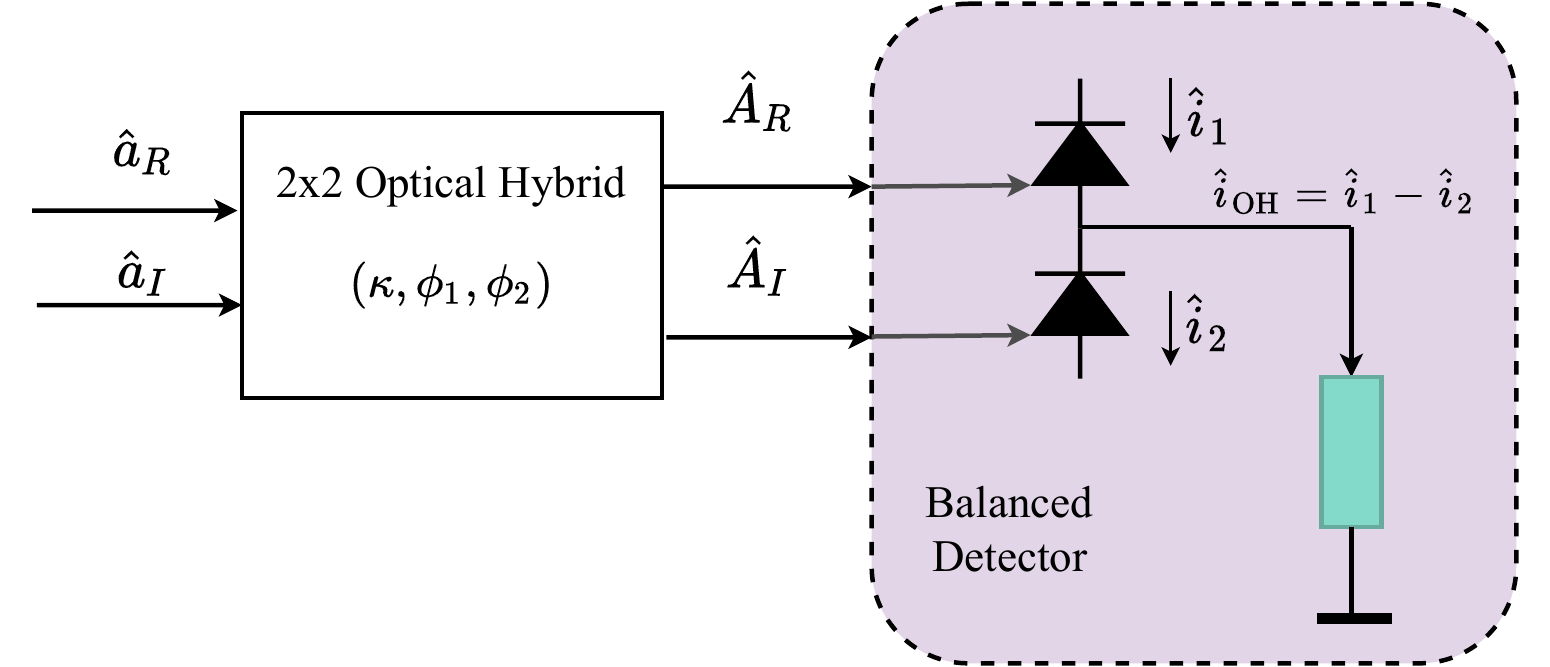}
\caption{Receiver configuration of a 2 × 2 optical hybrid-based joint balanced detection receiver.}
\label{fig:OH}
\end{figure}

In this section, we describe a practical receiver design using a 2x2 optical hybrid for EA communication. An optical hybrid-based joint detection scheme is suitable for EA communication as it can be directly implemented in integrated optics and quantum nanophotonics. For a two-dimensional constellation, a 2x2 optical hybrid receiver can be used, as shown in Figure~\ref{fig:OH}. A detailed discussion of the optical hybrid receiver design can be found in~\cite{djordjevic2021quantum}, where Gaussian modulation has also been discussed. The scattering matrix $\Scal$ of the 2x2 optical hybrid is described by Equation~\eqref{eq:OH_Scattering}
\spliteq
{
\label{eq:OH_Scattering}
\Scal = \m{ e^{j\phi_1}\sqrt{1-\kappa} &&  \sqrt{1-\kappa} \\ \sqrt{1-\kappa} &&  e^{j\phi_2}\sqrt{\kappa} }
}
where $\kappa$ is the power-splitting ratio of Y-junction in a 2x2 optical hybrid; and $\phi_1,$ and $\phi_2$ are phase shift parameters~\cite{guan2017compact}.
Return and idler at the receiver are transformed based on the scattering matrix given in Equation~\eqref{eq:OHtransform}.
\spliteq
{
\label{eq:OHtransform}
\m{\hat{A}_R \\ \hat{A}_I} = \Scal\m{\hat{a}_R \\ \hat{a}_I} .
}
We consider equal power splitting set by $\kappa = 0.5$ and  write the scattering matrix as
\spliteq
{
\label{eq:OHtransform2}
\m{\hat{A}_R \\ \hat{A}_I} = \cfrac{1}{\sqrt{2}}\m{e^{j\phi_1} && 1\\1 && e^{j\phi_2} }\m{\hat{a}_R \\ \hat{a}_I} 
}
For BPSK, $\ahat_R = \ahat_{R'}e^{j\theta}$ with $\theta \in \{0,\pi\}$. The photocurrent operator is given by 
\spliteq
{
\label{eq:OHEXP}
\ihat_{\text{OH}} & = \cfrac{1}{2}e^{-j\theta} ( e^{-j\phi_1} - e^{j\phi_2} ) \ahat_{R'}^\dagger \ahat_I + 
 \cfrac{1}{2}e^{j\theta} ( e^{j\phi_1} - e^{-j\phi_2} ) \ahat_I^\dagger \ahat_{R'} 
}
The expectation of photocurrent is given by 
\spliteq
{
\label{eq:exp_NOH}
N_{\text{OH}} & = \expval{\ihat_{\text{OH}}} = \cfrac{1}{2}e^{-j\theta}\sqrt{\eta N_s (N_s + 1)}( e^{-j\phi_1} - e^{j\phi_2} ) \\ & + \cfrac{1}{2}e^{j\theta}\sqrt{\eta N_s (N_s + 1)} ( e^{j\phi_1} - e^{-j\phi_2} )
}
In this paper, we consider a special case of 2x2 optical hybrid receiver where $\phi_1 = 0$ and $\phi_2 = \pi$, for which, $N_{\text{OH}} = 2\sqrt{\eta N_s (N_s + 1)}\cos\theta$.
The variance of the photocurrent operator is given by 
\spliteq
{
\label{eq:var_OH}
\sigma^2_\text{OH} & = \expval{\ihat_\OH^2} - \expval{\ihat_\OH}^2\\ 
& = \cfrac{1}{4}|e^{j\phi_1} -e^{-j\phi_2}|^2 (2 N_R N_I + N_R  + N_I )  \\ & + \cfrac{\eta N_s (N_s + 1)}{4}\bigg[ \bigg( \expval{e^{-2j\theta}} - e^{-2j\theta}\bigg) ( e^{-j\phi_1} - e^{j\phi_2} ) ^2\bigg] \\
& + \cfrac{\eta N_s (N_s + 1)}{4}\bigg[ \bigg( \expval{e^{2j\theta}} - e^{2j\theta}\bigg) ( e^{j\phi_1} - e^{-j\phi_2} ) ^2\bigg] \\ & -  \cfrac{\eta N_s (N_s + 1)}{2   } e^{-j\theta}e^{j\theta} | e^{j\phi_1} - e^{-j\phi_2}|^2
}
For equal prior BPSK symbols, $\expval{e^{\pm 2j\theta}} =  (e^{\pm 2j\cdot 0 } +e^{\pm 2j\cdot \pi})/2   = 1$. For non-equal prior symbols with priors $p_0$ and $p_1$, $\expval{e^{\pm 2j\theta}}$ is calculated as $p_0 e^{\pm 2j \cdot 0} +p_1e^{\pm 2j\cdot \pi}$ which is still 1. Further, regardless of phase value $\theta\in\{0,\pi\}$ for BPSK symbols, $e^{\pm 2j\theta} = \cos 2\theta$. Putting these values in Equation~\eqref{eq:var_OH}, and considering special case of $\phi_1 = 0$ and $\phi_2 = \pi$, the variance for BPSK is 
\spliteq
{
\label{eq:var_OH3}
\sigma^2_\text{OH}(\theta) & = (2 N_R N_I + N_R  + N_I )\\ & + 2\eta N_s (N_s + 1)(1-\cos2\theta) -2\eta N_s (N_s + 1)
}
where $N_R = \eta N_s + N_B$ and $N_I = N_s$. Similar to OPA and OPC receiver design, the decision scheme uses threshold detection for state discrimination. Since the target of this paper is a highly noisy and lossy environment, we choose $N_B = 1$, $N_s = 0.01$, $\eta = 0.01$, and $G = 1.1$ as a representative of such a condition.

\section{Evaluation of Entanglement-Assisted Communication Receivers}
\label{sec:evaluation}
\subsection{Error Probability Calculation}
\label{seC:pe_calculation}
The probability of error of state discrimination for the case of BPSK using OPA is given by
\spliteq{
\label{eq:BPSK_OPA_PE}
P_E & = p_0 P_\text{OPA}\bigg(n < \Nth|\theta=0; M,i\bigg) \\ & + p_1 \bigg[ 1 - P_\text{OPA}\bigg(n < \Nth|\theta=\pi; M, i\bigg) \bigg].
}
An optimum value of $\Nth$ can be found by equating individual error term in Equation~\eqref{eq:BPSK_OPA_PE} which, for case of symbols with equal priors gives us $\Nth(\theta) = \cfrac{M (\sigma(\pi)\Nbar(0) + \sigma(0)\Nbar(\pi))}{(\sigma(\pi) + \sigma(0))}$. Derivation of the optimum threshold for OPA is provided in Appendix~\ref{sec:opa_optimum_nth} that uses Gaussian approximation.

However, for unequal priors, the optimum threshold $\Nth$ is the one that satisfies the condition 
\spliteq{
\label{eq:optimum_threshold}
p_0 P_\text{OPA}&\bigg(n < \Nth|\theta=0; M, i\bigg) \\ & = p_1 \bigg[ 1 - P_\text{OPA}\bigg(n < \Nth|\theta=\pi; M, i\bigg)\bigg].
}
We solve Equation~\eqref{eq:optimum_threshold} for $\Nth$ using grid search procedure and plug into Equation~\eqref{eq:BPSK_OPA_PE} to calculate the error probability. In such a case, there is no closed-form solution.

\begin{figure}[htpb]
\centering
\includegraphics[width=0.8\linewidth, trim={1.5cm 3.5cm 1.5cm 3.5cm},clip]{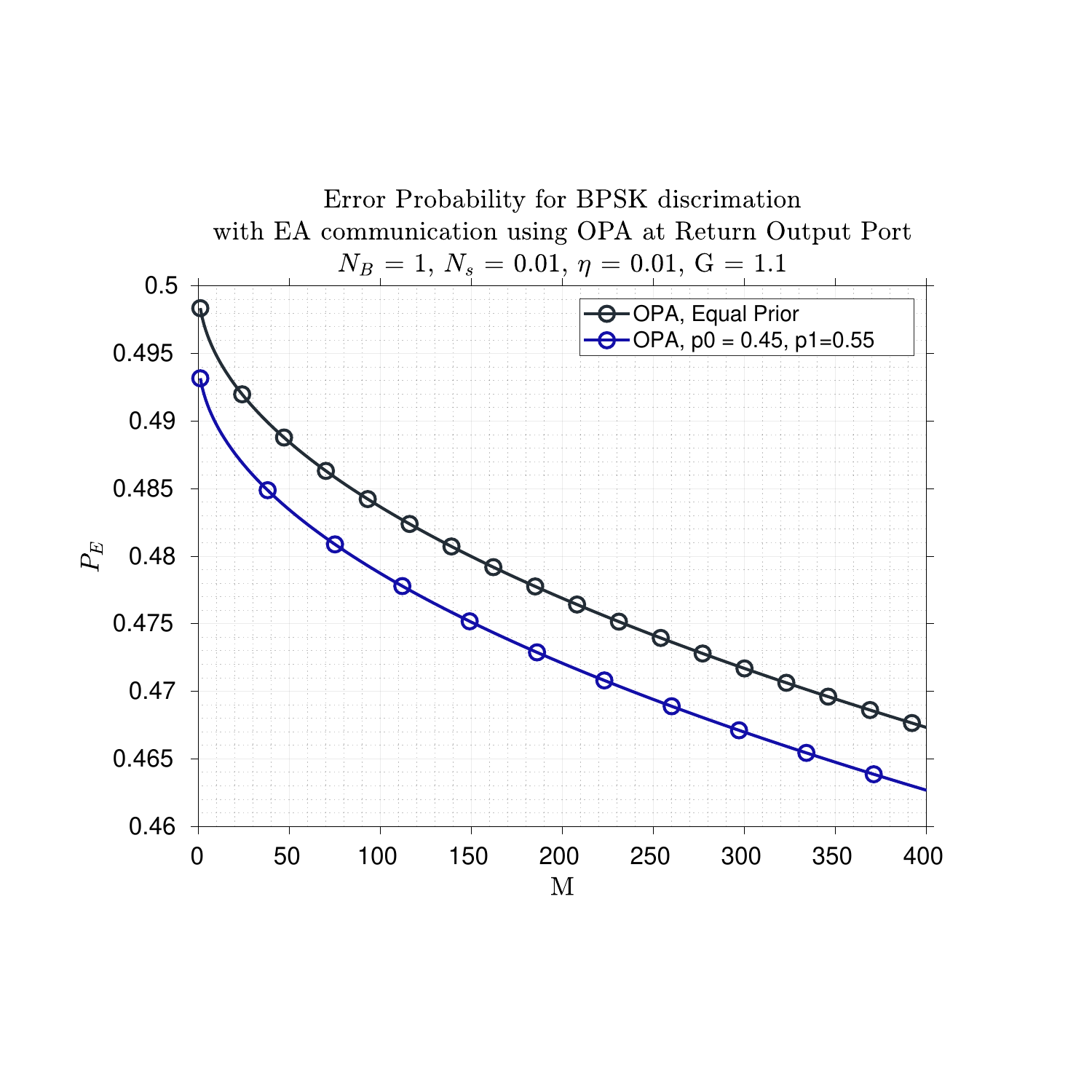}
\caption{\small{\textbf{Symbol-by-symbol (separable) minimum error-probability measurement on each return-idler mode pair at the return output port of the OPA receiver. As compared to measurement made at the idler output port of the OPA receiver, the error probability is higher at the return output port of the OPA receiver.}}}
\label{fig:PEPlot_Signal}
\end{figure}

\begin{figure}[htpb]
\centering
\includegraphics[width=0.75\linewidth, trim={0cm 0cm 0cm 0cm},clip]{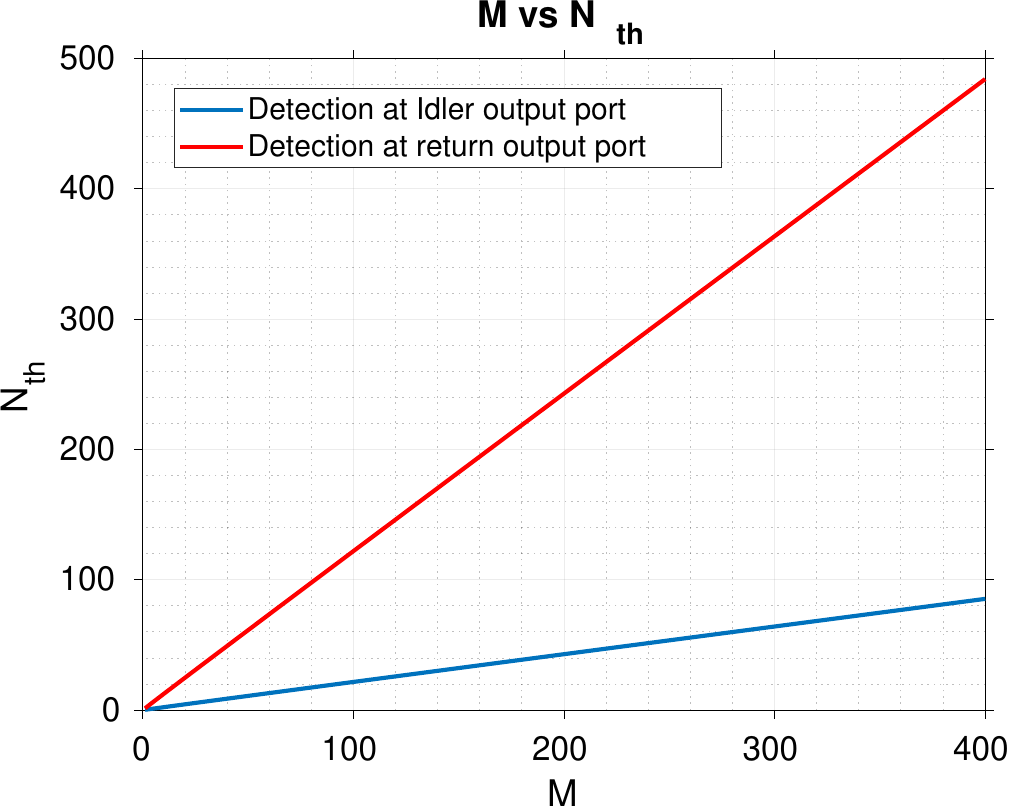}
\caption{Optimal threshold for hypothesis testing at the output port of OPA as a function of the number of modes. A higher threshold is required when detection is made at the return output port (red line) compared to the detection made at the idler output port of the OPA receiver (blue line). Unequal prior were $p_0 = 0.45, p_1= 0.55$. Plot was generated using $N_s = 0.01, N_B = 1, \eta = 0.01$.}
\label{fig:thresholing}
\end{figure}

\begin{figure}[htpb]
\centering
\includegraphics[width=0.99\linewidth, trim={0cm 1cm 0cm 0.2cm},clip]{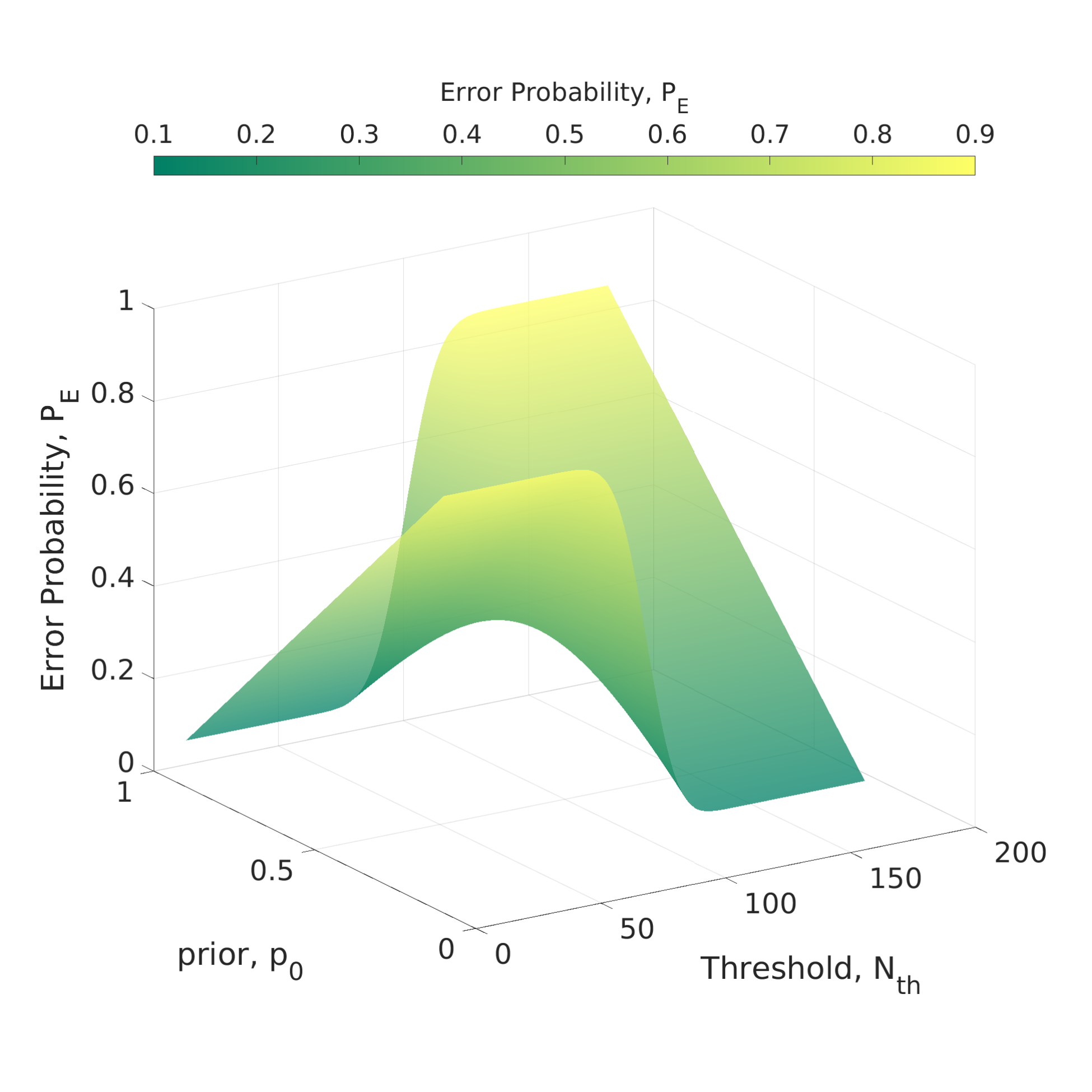}
\caption{\small{Surface plot of $P_E$ as a function of prior $p_0$ and threshold mean photon number $\Nth$ from Equation~\eqref{eq:BPSK_OPA_PE} for photodetection at idler output port for OPA receiver. The plot was generated using $N_s = 0.01, N_B = 1, \eta = 0.01$. These plots reveal that for a fixed threshold, error probability is maximum at equal prior while non-equal probable symbol seems to reduce error probability.}}
\label{fig:PE_Surfplot_a}
\end{figure}

\begin{figure}[htpb]
\centering
\includegraphics[width=0.99\linewidth, trim={0cm 0cm 0cm 0.2cm},clip]{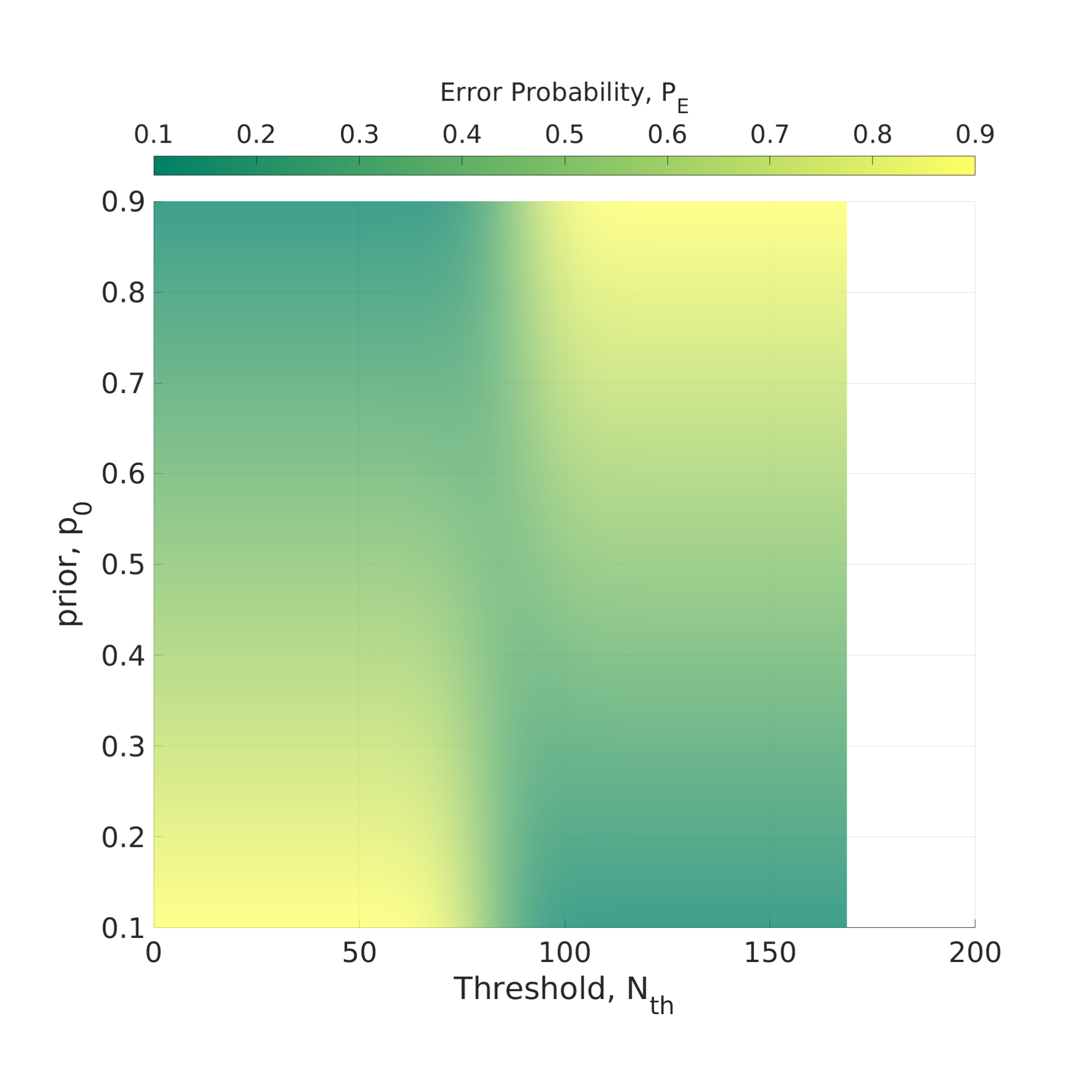}
\caption{\small{2D view of the surface plot from a alternative direction. Colormap denotes error probability -- darker color means lower error probability.}}
\label{fig:PE_Surfplot_b}
\end{figure}

\begin{figure}[htpb]
\centering
\includegraphics[width=0.99\linewidth, trim={0cm 0cm 0cm 0.2cm},clip]{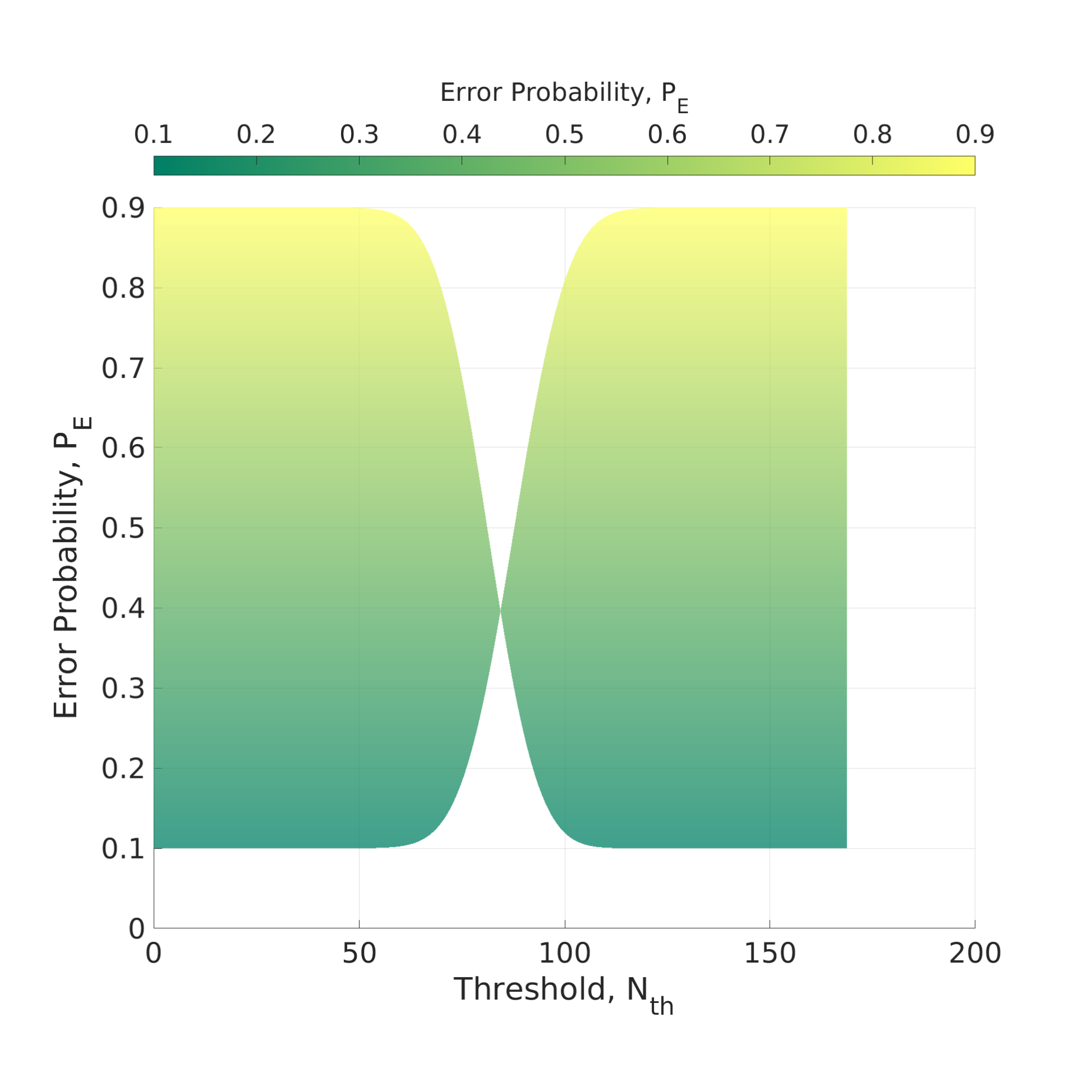}
\caption{\small{2D view of the surface plot from a alternative direction. Colormap denotes error probability -- darker color means lower error probability.}}
\label{fig:PE_Surfplot_c}
\end{figure}

\begin{figure}[htpb]
\centering
\includegraphics[width=0.99\linewidth, trim={0cm 0cm 0cm 0.2cm},clip]{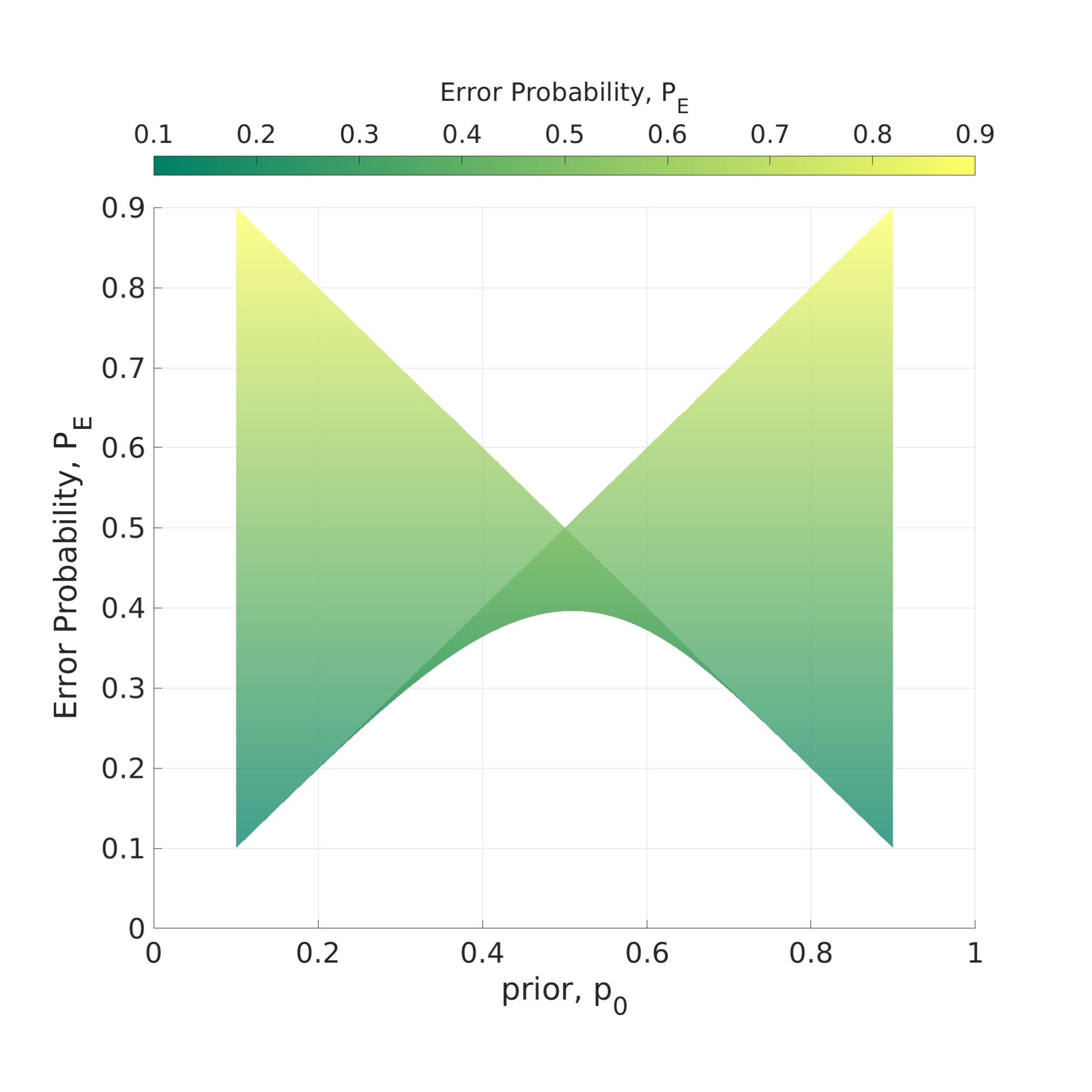}
\caption{\small{2D view of the surface plot from a alternative direction. Colormap denotes error probability -- darker color means lower error probability.}}
\label{fig:PE_Surfplot_d}
\end{figure}

The joint detection can be made either at the idler output port or the signal output port. The error probability of discrimination is higher at the return port compared to detection made at the idler port, as shown in Figure~\ref{fig:PEPlot_Signal}. As a result, our further analysis focuses solely on making joint detection at the idler port and \newchange{we drop the index $i$ from the probability notation moving forward}. We find that for the case of non-equal priors, the mean threshold photon number $\Nth$ for BPSK discrimination is higher for any detection made at the return output port than at the idler output port (see Figure~\ref{fig:thresholing}). Note that even though the error probability $P_E$ in Equation~\eqref{eq:BPSK_OPA_PE} is a convex function of $\Nth$, it is a monotonic function of the prior $p_0$, as shown in Figure~\ref{fig:PE_Surfplot_a}, Figure~\ref{fig:PE_Surfplot_b}, Figure~\ref{fig:PE_Surfplot_c}, and Figure~\ref{fig:PE_Surfplot_d}. Hence, there does not exist an optimum prior that minimizes the probability of error for state discrimination.

For the OPC receiver, we calculate the error probability by taking a Gaussian approximation of photodetection statistics similar to Equation~\eqref{eq:BPSK_OPA_PE}. This is because we measure the difference of photocurrent obtained at the two arms of the beamsplitter at the detection side (as shown in Figure~\ref{fig:OPC}). The Gaussian approximation yields the probability of error formula given in Equation~\eqref{eq:PE_OPC}.
{\small
\spliteq
{
\label{eq:PE_OPC}
P_E & = p_0 \Fcal_\OPC \bigg( \Nth, M\cdot N_\OPC(0), \sqrt{M}\cdot\sigma_\OPC(0)\bigg) \\ & + p_1 \bigg[ 1 - \Fcal_\OPC\bigg( \Nth, M\cdot N_\OPC(\pi), \sqrt{M}\cdot\sigma_\OPC(\pi)\bigg) \bigg]
}
}
In Equation~\eqref{eq:PE_OPC}, $N_\OPC(\theta)$ and $\sigma_\OPC$ are given by Equations \eqref{eq:OPCMean} and \eqref{eq:OPCVar} respectively. $\Fcal_\OPC$ is cumulative distribution function of a Gaussian distribution with mean $M\cdot N_\OPC(\theta)$ and standard deviation $\sqrt{M}\cdot\sigma_\OPC$. 

\begin{figure}[htpb]
\centering
\includegraphics[width=0.8\linewidth, trim={1.5cm 3.5cm 1.5cm 3.5cm},clip]{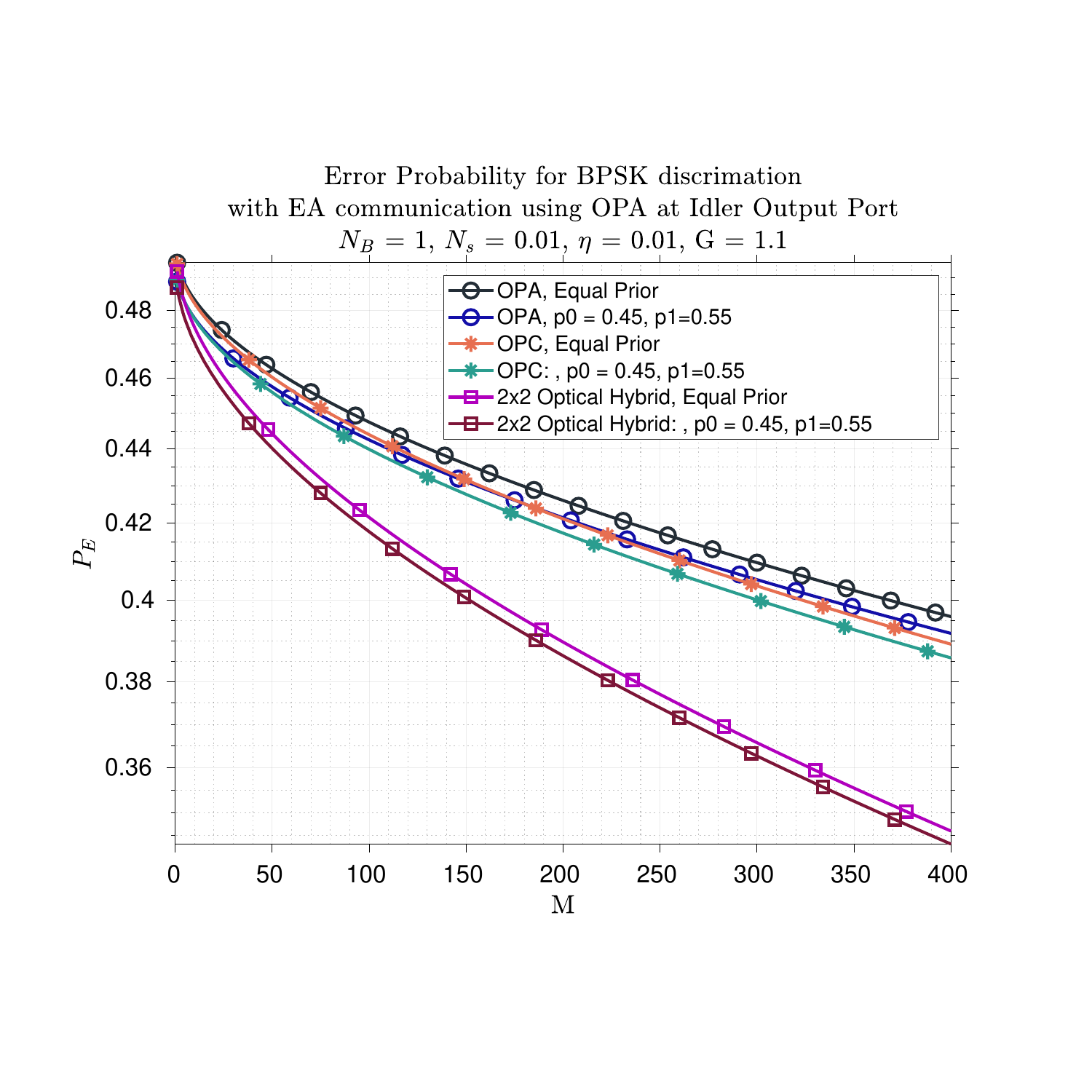}
\caption{\small{\textbf{Symbol-by-symbol (separable) minimum error-probability measurement on each return-idler mode pair at the idler output port. \changes{We find that unequal priors have a lower error probability as compared to BPSK symbols with equal prior.}}}}
\label{fig:PEPlot}
\end{figure}

Equation~\eqref{eq:PE_OPC} is similar to Equation~\eqref{eq:BPSK_OPA_PE} but written explicitly using the cumulative distribution function (CDF) notation. Like the OPA receiver, we can calculate the optimum $\Nth$ by equating the two terms of Equation~\eqref{eq:PE_OPC}. From Figure~\ref{fig:PEPlot}, we see that the OPC receiver's performance in terms of error probability in discriminating BPSK symbols is better than that of the OPA receiver. However, for a low number of modes $M$, OPA receivers with non-equal priors still perform better than OPC receivers with equal priors and perform similarly to OPC receivers with non-equal priors. Our evaluation suggests that lower-complexity receivers like OPA receivers with fewer optical components can provide superior information retrieval with a suitable choice of prior.

The error probability of a 2x2 optical hybrid can be calculated using a formula similar to the one in Equation~\eqref{eq:PE_OPC} with means and variances from Equations~\eqref{eq:exp_NOH} and \eqref{eq:var_OH3}, respectively. From the error probability plot in Figure~\ref{fig:PEPlot}, we see that the 2x2 optical hybrid offers a roughly $10\%$ improvement in BPSK state discrimination compared to the OPC receiver.

%\begin{figure}[htpb]
% \captionsetup[subfigure]{aboveskip=0pt,belowskip=1pt}
%\centering
%\begin{subfigure}{1.0\linewidth}
%\centering
%    \includegraphics[angle=0,origin=c,trim={1.0cm 7.0cm 2.0cm 7.0cm},clip,width=0.99\linewidth]{NegativeBinomial_Idler_Port_Optimal_Threshold.pdf}
%    \label{fig:idler_threshold}
%\end{subfigure}
%\begin{subfigure}{1.0\linewidth}
%    \centering
%    \includegraphics[angle=0,origin=c,trim={3.0cm 6.5cm 4.0cm 6.5cm},clip,width=0.99\linewidth]{NegativeBinomial_Signal_Port_Optimal_Threshold.pdf}
%    \label{fig:signal_threshold}
%\end{subfigure}
%\caption{Optimal threshold for hypothesis testing at the output port of OPA as a function of the number of modes. A higher threshold is required when detection is made at the return output port (bottom figure) compared to the detection made at the idler output port of the OPA receiver (top figure). Unequal prior were $p_0 = 0.45, p_1= 0.55$. Plot was generated using $N_s = 0.01, N_B = 1, \eta = 0.01$.}
%\label{fig:thresholing}
%\end{figure}

%%%%%%%%%%

\subsection{Mutual Information Calculation}
The Holevo capacity\cite{holevo2001evaluating, holevo2002entanglement} quantifies the maximum amount of information, in bits per channel use, that can be sent over a quantum channel when the use of entangled states at the input and arbitrary measurements at the output are permitted. In the situation under consideration in this work, \changes{the Holevo capacity evaluates to Equation}~\eqref{eq:holevocapacity}
\spliteq
{
\label{eq:holevocapacity}
C = g(\eta N_s + N_B) - g(N_B)
}
where 

\spliteq{
\label{eq:gfunc}
g(n) = (n+1)\log_2(n+1) - n\log_2 (n)
}
is the entropy of the thermal state with mean photon number $n$. To write the mutual information and, in turn, the capacity for entanglement-assisted classical communication that requires symbol-by-symbol joint detection, we are required to calculate the conditional probability distribution. Assuming that the random variable $X$ denotes the transmitted symbols and $Y$ denotes the detected symbols, we calculate the mutual information as follows: we first calculate the conditional probabilities to complete the transition matrix. Using the conditional probabilities, we can calculate the posteriors, which are then used to calculate the conditional entropies, followed by the calculation of the mutual information. The steps to calculate the mutual information are provided in Equation~\eqref{eq:mutual_information_calc} in Appendix~\ref{sec:app_mutual_info}.

\begin{figure}[htpb]
\centering
\includegraphics[width=0.94\linewidth, trim={0.0cm 0.0cm 0.0cm 0.0cm},clip]{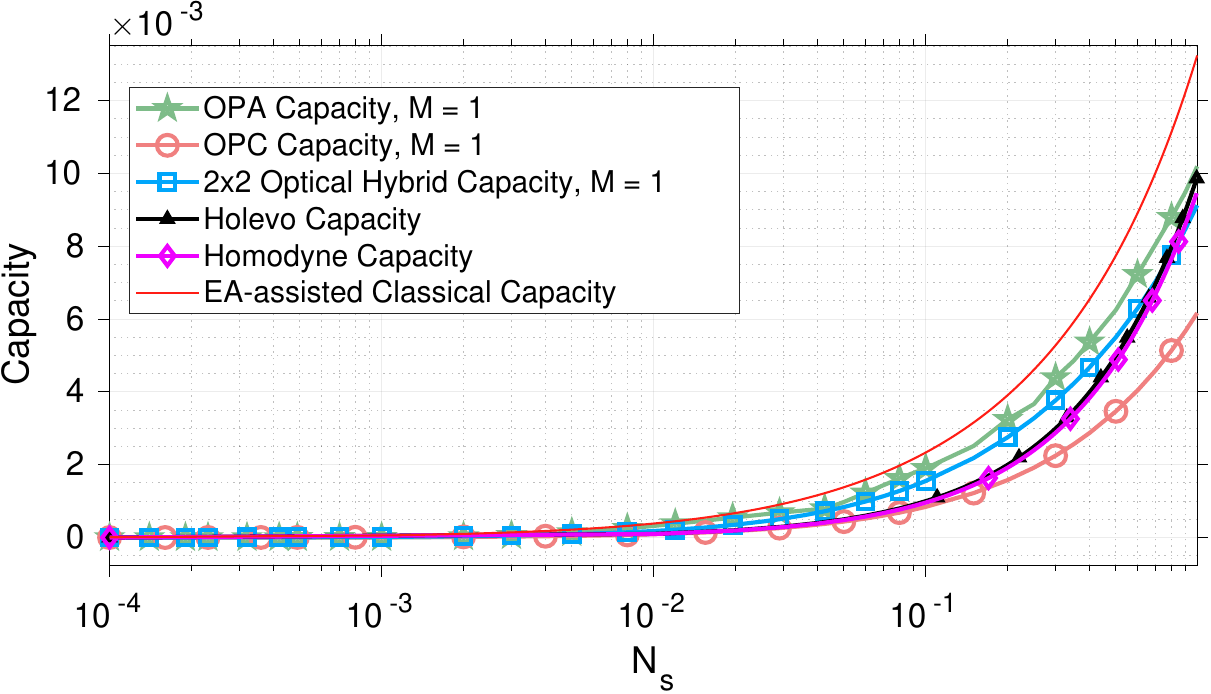}
\caption{Channel Capacity for BPSK state discrimination using different receiver schemes for single mode. We see that even for a single mode, entanglement assistance offers an advantage in terms of increasing the capacity of the channel and beats the classical capacities such as Holevo capacity and Homodyne Capacity. The plot was generated using $N_B = 1, \eta = 0.01$. The red line curve denotes the theoretical ultimate capacity that can be achieved through the entanglement-assistance.}
\label{fig:CapacityPlot_M1_1}
\end{figure}

\begin{figure}[htpb]
\centering
\includegraphics[width=0.47\linewidth, trim={0.0cm 0.0cm 0.0cm 0.0cm},clip]{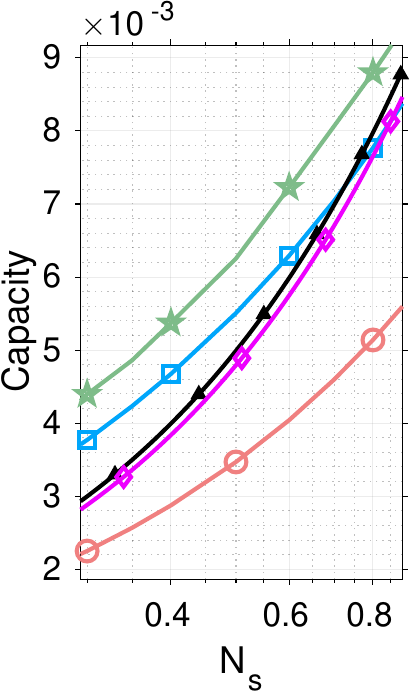}
\caption{Zoomed-in view to clearly illustrate that Holevo and Homodyne capacity are not exactly the same. The plot was generated using $N_B = 1, \eta = 0.01$.}
\label{fig:Capacity_plot_M_1_zoomed}
\end{figure}

\begin{figure}[htpb]
\centering
\includegraphics[width=1.0\linewidth, trim={0.0cm 0.0cm 0.0cm 0.0cm},clip]{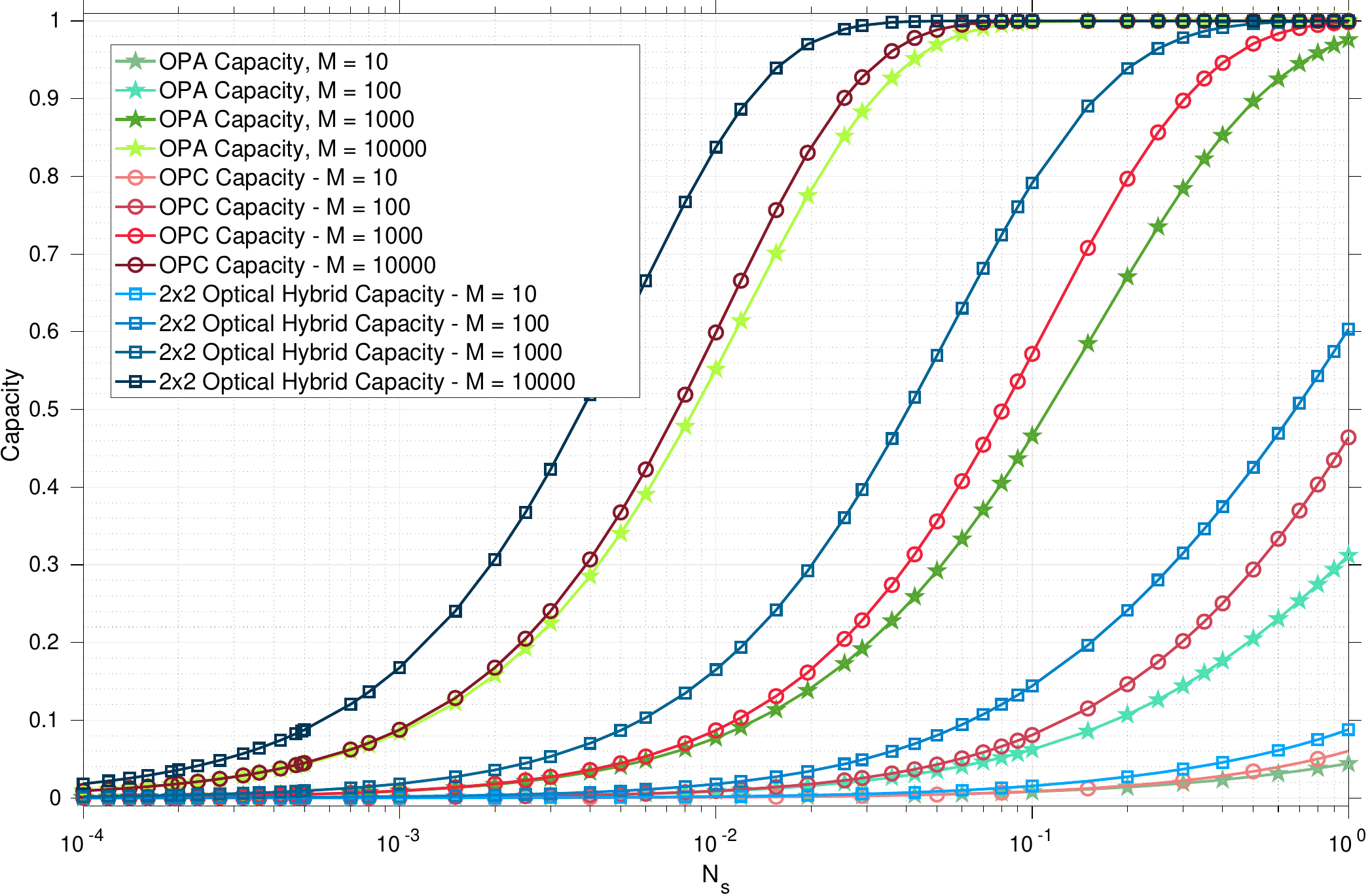}
\caption{Channel Capacity for BPSK state discrimination using different receiver schemes. We see that entanglement assistance offers an advantage in terms of increasing the capacity of the channel. We chose the transmittivity of the Bosonic channel as $\eta = 0.01$ and the mean background photon as $N_B = 1.0$. In our numerical study, classical capacity stays below 0.07 bits per channel use for mean signal photon $N_s < 1.0$.}
\label{fig:CapacityPlot}
\end{figure}

The Shannon's capacity for transmitting classical information with our EA receivers can be calculated by taking the maximum of mutual information over prior $p$ and threshold mean photon number $\Nth$, i.e.
\spliteq
{
\label{eq:CapacityEQ}
C_{\text{EA}} = \max_{p, \Nth} I(X;Y)
}
We find that symbols with equal priors maximize the mutual information, as expected. We conducted a simulation study with a varying number of modes $M$ to optimize the mutual information as a function of the signal mean photon number transmitted over a noisy Bosonic channel with $N_B = 1$ and transmittivity $\eta = 0.01$. In \newchange{Figure}~\ref{fig:CapacityPlot_M1_1} and \newchange{Figure}~\ref{fig:CapacityPlot}, \newchange{we present a comparison of the capacities of various receiver designs proposed for the BPSK} constellation using Equation~\ref{eq:CapacityEQ}. At the same time, we also plot the capacity of the Homodyne receiver, where the average number of photons received is $4\eta N_s$ and the average number of noisy photons is $2N_B + 1$. The capacity of a Homodyne receiver is given by $C_H = 0.5\log_2[1 + 4\eta N_s/(2N_B + 1)]$~\cite{shapiro2005ultimate, takeoka2014capacity}. As a reference, we also plot the Holevo capacity, given by Equation~\eqref{eq:holevocapacity}. Note that the Holevo capacity requires coherent states with Gaussian modulation. \newchange{For EA communication with the BPSK constellation, we find that joint receivers based on OPA and OH proposed in this paper outperform the Holevo capacity, even for a single mode}, as shown in Figure~\ref{fig:CapacityPlot_M1_1} and Figure~\ref{fig:Capacity_plot_M_1_zoomed}. We also conducted simulation studies for a large number of modes. From our analysis and results, shown in Figure~\ref{fig:CapacityPlot}, we conclude that for the proposed EA receiver design employing a joint-detection scheme, a large number of signal-idler modes is not required. For $M=1$, the OPA receiver's performance is better than the OPC receiver and 2x2 optical hybrid receiver. \newchange{C-band operates at $35nm$, hence $d\lambda  = 35nm$.  The central wavelength, $\lambda = 1550 nm$. The typical observation interval of symbols is $1\mu s$. Phase matching bandwidth of the C-band is calculated as} 

\spliteq
{
\label{eq:phbw}
B = \cfrac{c}{\lambda^2}d\lambda = \cfrac{3\times 10^8}{(1550\times 10^{-9})^2} \cdot 35\times 10^{-9} = 4.3704 \times 10^{12}
}
\newchange{Then the number of bosonic mode, $M$ for C-band is $4.3704 \times 10^{12} \cdot 1 \mu s = 4.3704 \times 10^{6}$. Hence if we were to use the number of modes in the order of $10^6$, we would end up using C-band as was demonstrated in} \cite{shi2020practical, hao2021entanglement}. \newchange{However, as we have shown above, achieving a superior performance requires as little as $ M = 1$ if we choose optimal values of prior and threshold mean photon.}

We also compared our capacities with ultimate bound, i.e., entanglement-assisted classical capacity $C_{\text{ultimate}}$ as described in~\cite{bennett2002entanglement, djordjevic2021entanglement2}. To calculate the $C_{\text{ultimate}}$, we adopted Equation~\eqref{eq:cultimate} from~\cite{bennett2002entanglement, djordjevic2021entanglement2} with suitable modifications as per the use case described in Section~\ref{sec:receiver_design}. Entanglement-assisted classical capacity is given by
\begin{equation}
\label{eq:cultimate}
\begin{split}
C_{\text{ultimate}} = g(N_s) + g(N_R) -  \bigg( g( \tfrac{\nu_+ - 1}{2} )  + g( \tfrac{\nu_- - 1}{2} ) \bigg)
\end{split}
\end{equation}
with $ a = 2N_s + 1$,  $b = 2N_R + 1$, 
$C_\eta = 2\sqrt{\eta N_s (N_s +1)}$,
$\nu_\pm = \bigg[\sqrt{(a+b)^2 - 4C_\eta^2} \pm (b - a)\bigg]/2$,
and $g()$ has been defined in Equation~\eqref{eq:gfunc}. From Figure~\ref{fig:CapacityPlot_M1_1}, it is evident that further development in the receiver design is required to achieve performance closer to the ultimate bound of the capacity given by Equation~\eqref{eq:cultimate}.

\begin{figure}[htpb]
\centering
\includegraphics[width=1.0\linewidth, trim={2.4cm 1.0cm 2.4cm 1.0cm},clip]{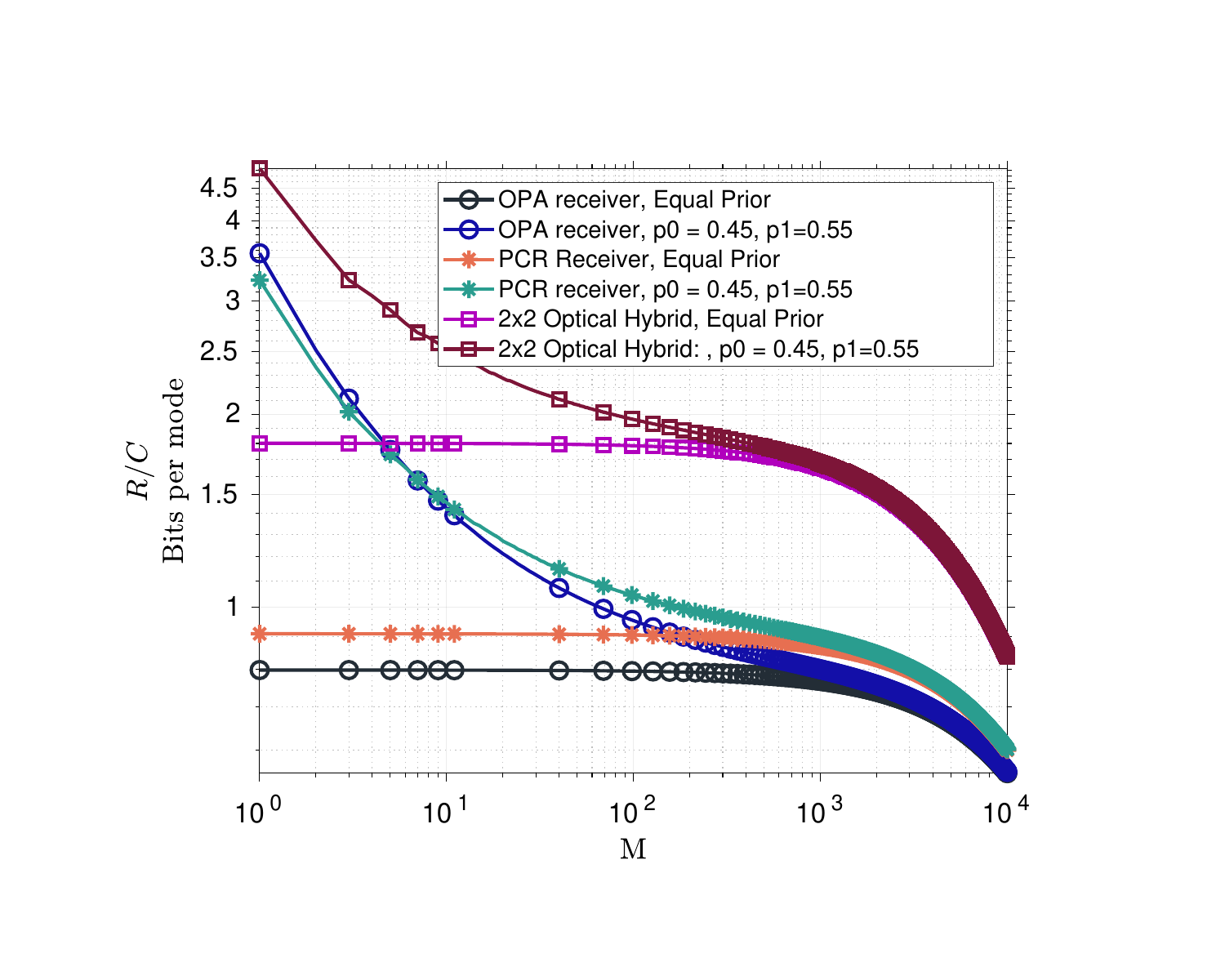}
\caption{Information rate as a function of the number of modes $M$ for OPA and OPC receiver. 
It is evident that as we increase the number of modes the $R/C$ for equal and unequal priors attain the same values. Further, we see better performance in terms of information rate when the number of modes is low. The plot was generated using $N_s = 0.01, N_B = 1, \eta = 0.01$.}
\label{fig:commrate}
\end{figure}

For a large number of modes, the 2x2 optical hybrid receiver performs better in terms of capacity as shown in Figure~\ref{fig:CapacityPlot}. \newchange{Previous work}~\cite{shapiro2005ultimate, takeoka2014capacity} \newchange{in this direction have not considered the use of OH receivers. An OH receiver uses a balanced detector, similar to OPA to distinguish the modulation which is more practical and efficient. Further, OH is known to suppress noise as discussed in}~\cite{SHIEH2009263}. It could be argued that the capacity should be divided by the number of modes for a scheme using multiple modes. However, in this paper, we are talking about the overall receiver design's capacity, rather than bits per mode. In addition, we find that a number of modes greater than 1 may not be needed to outperform the Holevo capacity, as seen from Figure~\ref{fig:CapacityPlot_M1_1}. \newchange{Thus, for a well-designed receiver, repetition coding may not always be useful. This claim is further corroborated by Figure}~\ref{fig:commrate}.

\begin{figure}[htpb]
\centering

\end{figure}

Additionally, we also plot the per-mode communication rate $R$, normalized by the Holevo capacity for classical communication $C$, in Figure~\ref{fig:commrate}. The communication rate $R$ is given by Equation~\eqref{eq:commrate}.
\spliteq
{
\label{eq:commrate}
R = \cfrac{1 + P_e\log_2(P_e) + ( 1 - P_e)\log_2(1 - P_e)}{M}.
}
Equation~\eqref{eq:commrate} is based on symmetric hypothesis testing. We find that in terms of the normalized communication rate, the OPA and OPC receivers perform almost three times better in the photon-starved regime when BPSK symbols with non-equal priors are used, compared to when BPSK symbols are equally likely. At the same time, the 2x2 optical hybrid receiver for non-equal priors performs roughly 2.5 times better, compared to BPSK with equal priors in the photon-starved regime. \changes{Furthermore, the 2x2 optical hybrid receiver can outperform an OPA-based receiver by as much as 30\% in terms of information rate. Finally, it's worth noting that a large number of modes does not necessarily equate to superior performance.}

\subsection{Discussion}
\newchange{It is expected that the order of error probability of two receiver designs should be the reverse of the order of capacity when compared. We find that this is not the case for OPA, OPC, and OH as evident from Figure}~\ref{fig:PEPlot} as well as Figure~\ref{fig:CapacityPlot_M1_1} and Figure~\ref{fig:CapacityPlot}. \newchange{The ordering in the error-probability plot and capacity plot is only related when mutual information is optimized with respect to the prior only. However, we optimize mutual information with respect to the prior as well as threshold mean photon number as shown in Equation}~\ref{eq:CapacityEQ}. Hence, the usual ordering relationship is no longer applicable. Further, as the number of modes increases, the probability distribution seen in Equation~\ref{eq:OPA_P} resembles more and more Gaussian. \newchange{Thus the ordering of capacity is altered as we move from the number of modes $M = 1$ to higher modes. Thus, with the higher number of modes, the OH receiver design has the best performance, followed by OPC and then the OPA receiver. However, with $M=1$, the best performance is obtained by using the OPA receiver, followed by the OH receiver and then the OPC receiver.}

\section{Concluding Remarks and Future Works}
\label{eq:conclusion}
Entanglement is a unique phenomenon in quantum information science that can be leveraged to design new types of sensors, allowing computing devices to solve problems that are intractable for conventional computers. In communication systems, the use of entanglement assistance offers a unique advantage in terms of providing a better communication rate in low-photon number regimes. Pre-shared entanglement can be used to surpass the performance of classical capacities and the Holevo capacity in highly noisy and low-brightness conditions. However, there are several challenges in terms of the practical realization of entanglement, such as: (i) transmitting entanglement over long distances is challenging, and (ii) the optimum quantum receiver to achieve entanglement-assisted channel capacity has not yet been derived. Nevertheless, simulation results indicate that even when entanglement is not perfect, entanglement-assisted (EA) communication based on signal-idler pairs outperforms the Holevo capacity and the capacities of classical channels.

In this paper, we analyze several low-complexity receiver designs employing optical hybrids and balanced detectors. We demonstrate that for BPSK modulation, a 2x2 optical hybrid-based joint detection can outperform the OPA and optical phase-conjugation receivers. \changes{Numerical results demonstrate that we do not need a large number of signal-idler modes to outperform the Holevo and Homodyne capacities.}

\section{Disclosure}
The authors declare no conflicts of interest.
%\bibliographystyle{IEEEtran}

%\onecolumn

\section*{Appendix}

\section{Gaussian Approximation to Negative Binomial Photon Statistics}
\label{sec:opa_optimum_nth}
Although the photodetection statistics given by OPA receivers in Equations~\eqref{eq:OPA_P} are of negative binomial nature, they can be computationally expensive to calculate for large values of $M$. By recognizing that Equation~\eqref{eq:BPSK_OPA_PE} contains cumulative distributions and approximating them as Gaussian distributions, we can rewrite the equation as the error function (erf), which is commonly used to write the cumulative distribution function of a Gaussian distribution:

\begin{widetext}
{
\spliteq
{
\cfrac{1}{2}\bigg(1 + \textrm{erf}\bigg(\cfrac{\Nth - M \cdot \Nbar(0)}{\sqrt{2}\sqrt{M}\sigma(0)}\bigg)\bigg) & = 1 - \cfrac{1}{2}\bigg(1 + \textrm{erf}\bigg(\cfrac{\Nth - M \cdot \Nbar(\pi)}{\sqrt{2}\sqrt{M}\sigma(\pi)}\bigg)\bigg) \\
\Rightarrow \cfrac{1}{2} +\cfrac{1}{2} \textrm{erf}\bigg(\cfrac{\Nth - M \cdot \Nbar(0)}{\sqrt{2}\sqrt{M}\sigma(0)}\bigg)  & = \cfrac{1}{2} - \cfrac{1}{2} \textrm{erf}\bigg(\cfrac{\Nth - M \cdot \Nbar(\pi)}{\sqrt{2}\sqrt{M}\sigma(\pi)}\bigg)\\
 \Rightarrow \textrm{erf}\bigg(\cfrac{\Nth - M \cdot \Nbar(0)}{\sqrt{2}\sqrt{M}\sigma(0)}\bigg) & = - \textrm{erf}\bigg(\cfrac{\Nth - M \cdot \Nbar(\pi)}{\sqrt{2}\sqrt{M}\sigma(\pi)}\bigg)\\
 \textrm{Considering that erf}(-x)  = -\textrm{erf}(x) \\\textrm{and equating the arguments of erf}\\
 \Nth(\theta)  = \cfrac{M (\sigma(\pi)\Nbar(0) + \sigma(0)\Nbar(\pi))}{(\sigma(\pi) + \sigma(0))}.
}
}
\end{widetext}

However, we should be aware of how we may misinterpret the true performance of receivers due to approximation.

\begin{figure}[htpb]
\centering
\includegraphics[width=1.0\linewidth, trim={2.6cm 0.6cm 3.0cm 0.8cm},clip]{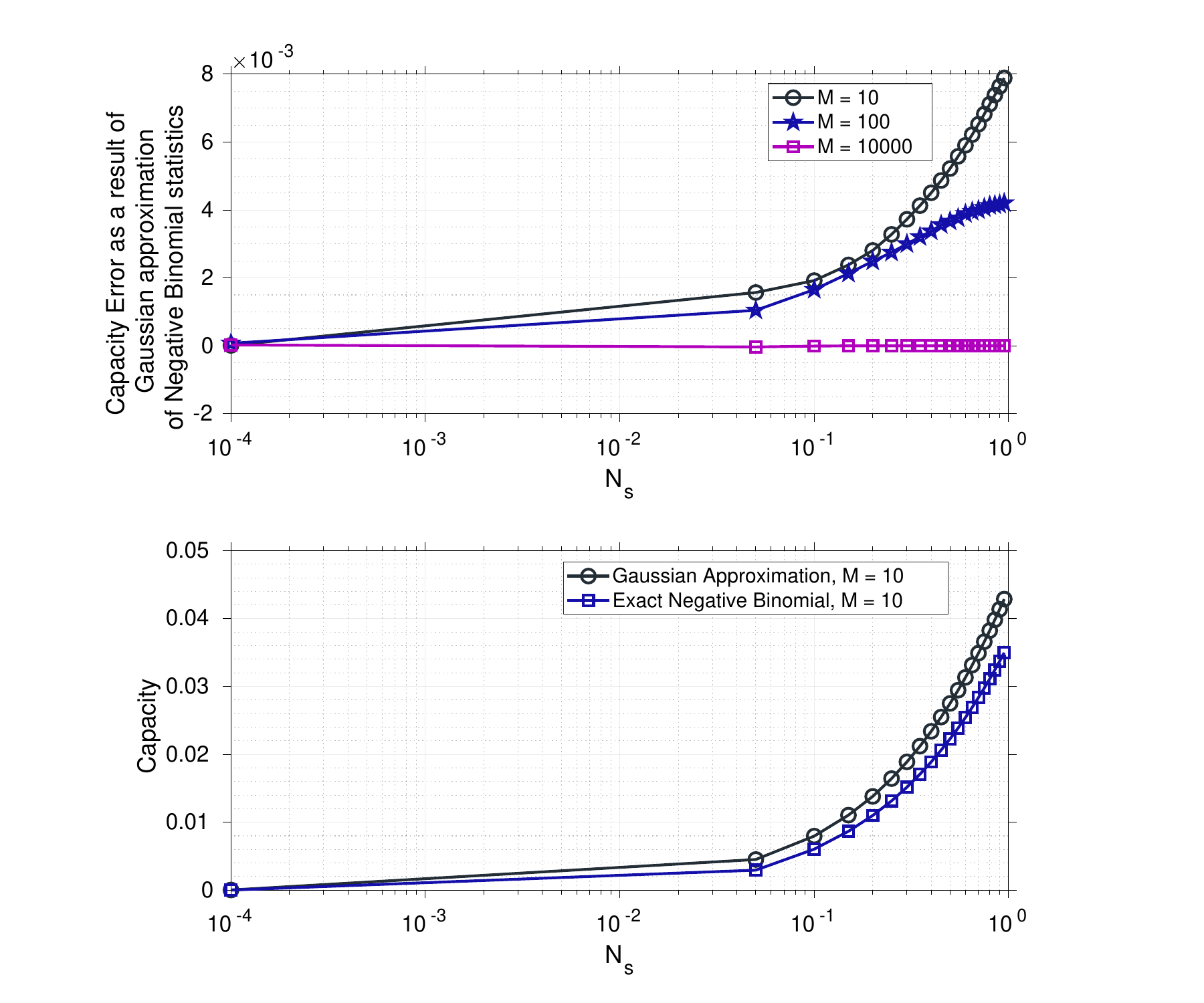}
\caption{\textbf{Top:} Capacity error in the case of OPA receiver as a result of approximating negative binomial photodetection statistics to Gaussian. \textbf{Bottom:} Capacity of OPA receiver with the number of modes $M=10$. We observe that with Gaussian approximation, we overestimate Shannon's capacity of the OPA receiver for EA communication while discriminating against BPSK states.}
\label{fig:GaussVSNB}
\end{figure}

In Figure~\ref{fig:GaussVSNB}, we plot the difference between $C_{\text{Gaussian}}$ and $C_{\text{NB}}$. $C_{\text{Gaussian}}$ represents the capacity of the OPA receiver as discussed in Section~\ref{sec:OPA}, where the photodetection statistics are approximated as Gaussian. $C_{\text{NB}}$ represents the capacity using the exact negative binomial distribution from Equation~\eqref{eq:OPA_P}. Our calculations have led us to the following observations: (i) the error of the approximation increases as the signal mean photon number, $N_s$, increases; (ii) with the Gaussian approximation, the capacity of the channel is overestimated compared to its true value; (iii) as the number of modes increases, the error of the approximation decreases. These observations are depicted in Figure~\ref{fig:GaussVSNB}. Although the Gaussian approximation overestimates the capacity, the error is of the order of $10^{-3}$, which is small compared to the value of the capacity and enables faster numerical calculations.

\begin{widetext}
\section{Derivation of Mean photon number for Optical Parametric Amplifier}
\label{sec:opa_mean_derivation}
\newchange{In this section, we derive in detail, the mean photon number for OPA using Equations} from Section~\ref{sec:OPA}.
{
\spliteq{
\label{eq:OPA_meanphotonOutput_A}
\uhat^\dagger \uhat & =( \sqrt{G}\ahat_R ^\dagger + \sqrt{G-1}\ahat_I) ( \sqrt{G}\ahat_R + \sqrt{G-1}\ahat_I^\dagger )\\[4pt]
 & =G\ahat_R^\dagger\ahat_R  + \sqrt{G(G-1)}\ahat_R^\dagger\ahat_I^\dagger  + \sqrt{G(G-1)}\ahat_I\ahat_R +  (G-1)\ahat_I\ahat_I^\dagger\\[4pt]
& =G\ahat_R^\dagger\ahat_R  + (\sqrt{G(G-1)})(\ahat_R^\dagger\ahat_I^\dagger + \ahat_I\ahat_R)  +  (G-1)\ahat_I\ahat_I^\dagger\\[6pt]
\expval{\uhat^\dagger \uhat} & = G\expval{\ahat_R^\dagger\ahat_R}  + (\sqrt{G(G-1)})(\expval{\ahat_R^\dagger\ahat_I^\dagger} + \expval{\ahat_I\ahat_R}) +  (G-1)\expval{\ahat_I\ahat_I^\dagger}\\[4pt]
& = GN_R + \sqrt{G(G-1)}(e^{j\theta} +e^{-j\theta} )\sqrt{\eta N_s(N_s + 1)}  + (G-1) ( 1 + N_I)\\[4pt]
& \text{As,~} \ahat_R = \ahat_{R'}e^{j\theta},~\expval{\ahat_{R'} \ahat_I} = \sqrt{\eta N_s (N_s + 1)}, \\ & ~\ahat_I \ahat_I^\dagger = \ahat_I^\dagger \ahat_I + I, \text{\newchange{(from the commutative property of annihilation and creation operators)}}\\[4pt]
& \text{Further, } N_R =  \eta N_s +N_B~\text{after passing through a channel}\\ & \text{~~~~with mean thermal photon number $N_B$}\\[4pt]
& N_I = N_s, ~\text{As,} \text{~~idler is per-shared} \\[4pt]
\Nbar_1(\theta) = \expval{\uhat^\dagger \uhat} & = G(\eta N_s + N_B) + (G-1)(1 + N_s) \\ & + 2\cos\theta \sqrt{G(G-1)}\sqrt{\eta N_s (N_s + 1)}
}
}

Similarly, 
{
\spliteq{
\label{eq:OPA_meanphotonOutput2_A}
\vhat^\dagger \vhat & =  ( \sqrt{G}\ahat_I^\dagger + \sqrt{G-1}\ahat_R)(\sqrt{G}\ahat_I + \sqrt{G-1}\ahat_R^\dagger)\\[4pt]
\expval{\vhat^\dagger \vhat} & = GN_I + 2\cos\theta \sqrt{G(G-1)}\sqrt{\eta N_s (N_s + 1)}  +  (G-1) ( 1 + N_R)\\[4pt]
\Nbar_2(\theta)= \expval{\vhat^\dagger \vhat} & =  GN_s + (G-1)( 1 + \eta N_s + N_B) + 2\cos\theta \sqrt{G(G-1)}\sqrt{\eta N_s (N_s + 1)} 
}
}

\end{widetext}

\section{Mutual Information Calculation}
\label{sec:app_mutual_info}
\newchange{In this section, at a very high level,  we provide a calculation of how mutual information can be calculated. The values of $P_\text{OPA}$ can be plugged from Equation}~\ref{eq:OPA_P}.
\newchange{We can first write the conditional probabilities, assuming that the random variable $X$ denotes the transmitted symbols and $Y$ denotes the detected symbols:}

\spliteq
{
\label{eq:condition_prob}
p_{y|x}(Y= 0 | X = 0) & = 1 -  P_\text{OPA}(n < \Nth|\theta = 0; M)\\
p_{y|x}(Y= 1 | X = 1) & = P_\text{OPA}(n < \Nth|\theta = \pi; M)\\
p_{y|x}(Y= 0 | X = 1) & = 1- P_\text{OPA}(n < \Nth|\theta = \pi; M)\\
p_{y|x}(Y= 1 | X = 0) & =  P_\text{OPA}(n < \Nth|\theta = 0; M)\\
p_y(Y=0) & = p_0 p_{y|x}(Y= 0 | X = 0)  + p_1 p_{y|x}(Y= 0 | X = 1) \\
p_y(Y=1) & = p_0 p_{y|x}(Y= 1 | X = 0)  + p_1 p_{y|x}(Y= 1 | X = 1) }

\newchange{Using conditional probabilities, we can obtain mutual information as follows}.
\spliteq
{
\label{eq:mutual_information_calc}
H(Y|X = 0) & = - p_{y|x}(Y= 0 | X = 0) \log_2(p_{y|x}(Y= 0 | X = 0)) \\ & - p_{y|x}(Y= 1 | X = 0) \log_2( p_{y|x}(Y= 1 | X = 0))\\
H(Y|X = 1) & = - p_{y|x}(Y= 0 | X = 1) \log_2( p_{y|x}(Y= 0 | X = 1) ) \\ & - p_{y|x}(Y= 1 | X = 1) \log_2(p_{y|x}(Y= 1 | X = 1))\\
H(Y|X) & = p_0 H(Y|X = 0) + p_1 H(Y|X = 1)\\
H(Y) &=  - p_y(Y=0) \log_2(p_y(Y=0) )\\ & - p_y(Y=1)\log_2(p_y(Y=1))\\
I(X;Y) &= H(Y) - H(Y|X)
}
\newchange{The mutual information can be used to calculate the capacity using Equation}~\ref{eq:CapacityEQ}.

\bibliography{apssamp}

\begin{thebibliography}{30}%
\makeatletter
\providecommand \@ifxundefined [1]{%
 \@ifx{#1\undefined}
}%
\providecommand \@ifnum [1]{%
 \ifnum #1\expandafter \@firstoftwo
 \else \expandafter \@secondoftwo
 \fi
}%
\providecommand \@ifx [1]{%
 \ifx #1\expandafter \@firstoftwo
 \else \expandafter \@secondoftwo
 \fi
}%
\providecommand \natexlab [1]{#1}%
\providecommand \enquote  [1]{``#1''}%
\providecommand \bibnamefont  [1]{#1}%
\providecommand \bibfnamefont [1]{#1}%
\providecommand \citenamefont [1]{#1}%
\providecommand \href@noop [0]{\@secondoftwo}%
\providecommand \href [0]{\begingroup \@sanitize@url \@href}%
\providecommand \@href[1]{\@@startlink{#1}\@@href}%
\providecommand \@@href[1]{\endgroup#1\@@endlink}%
\providecommand \@sanitize@url [0]{\catcode `\\12\catcode `\$12\catcode
  `\&12\catcode `\#12\catcode `\^12\catcode `\_12\catcode `\%12\relax}%
\providecommand \@@startlink[1]{}%
\providecommand \@@endlink[0]{}%
\providecommand \url  [0]{\begingroup\@sanitize@url \@url }%
\providecommand \@url [1]{\endgroup\@href {#1}{\urlprefix }}%
\providecommand \urlprefix  [0]{URL }%
\providecommand \Eprint [0]{\href }%
\providecommand \doibase [0]{http://dx.doi.org/}%
\providecommand \selectlanguage [0]{\@gobble}%
\providecommand \bibinfo  [0]{\@secondoftwo}%
\providecommand \bibfield  [0]{\@secondoftwo}%
\providecommand \translation [1]{[#1]}%
\providecommand \BibitemOpen [0]{}%
\providecommand \bibitemStop [0]{}%
\providecommand \bibitemNoStop [0]{.\EOS\space}%
\providecommand \EOS [0]{\spacefactor3000\relax}%
\providecommand \BibitemShut  [1]{\csname bibitem#1\endcsname}%
\let\auto@bib@innerbib\@empty
%</preamble>
\bibitem [{\citenamefont {Holevo}\ and\ \citenamefont
  {Werner}(2001)}]{holevo2001evaluating}%
  \BibitemOpen
  \bibfield  {author} {\bibinfo {author} {\bibfnamefont {A.~S.}\ \bibnamefont
  {Holevo}}\ and\ \bibinfo {author} {\bibfnamefont {R.~F.}\ \bibnamefont
  {Werner}},\ }\href@noop {} {\bibfield  {journal} {\bibinfo  {journal}
  {Physical Review A}\ }\textbf {\bibinfo {volume} {63}},\ \bibinfo {pages}
  {032312} (\bibinfo {year} {2001})}\BibitemShut {NoStop}%
\bibitem [{\citenamefont {Holevo}(2002)}]{holevo2002entanglement}%
  \BibitemOpen
  \bibfield  {author} {\bibinfo {author} {\bibfnamefont {A.~S.}\ \bibnamefont
  {Holevo}},\ }\href@noop {} {\bibfield  {journal} {\bibinfo  {journal}
  {Journal of Mathematical Physics}\ }\textbf {\bibinfo {volume} {43}},\
  \bibinfo {pages} {4326} (\bibinfo {year} {2002})}\BibitemShut {NoStop}%
\bibitem [{\citenamefont {Bennett}\ \emph {et~al.}(2002)\citenamefont
  {Bennett}, \citenamefont {Shor}, \citenamefont {Smolin},\ and\ \citenamefont
  {Thapliyal}}]{bennett2002entanglement}%
  \BibitemOpen
  \bibfield  {author} {\bibinfo {author} {\bibfnamefont {C.~H.}\ \bibnamefont
  {Bennett}}, \bibinfo {author} {\bibfnamefont {P.~W.}\ \bibnamefont {Shor}},
  \bibinfo {author} {\bibfnamefont {J.~A.}\ \bibnamefont {Smolin}}, \ and\
  \bibinfo {author} {\bibfnamefont {A.~V.}\ \bibnamefont {Thapliyal}},\
  }\href@noop {} {\bibfield  {journal} {\bibinfo  {journal} {IEEE Transactions
  on Information Theory}\ }\textbf {\bibinfo {volume} {48}},\ \bibinfo {pages}
  {2637} (\bibinfo {year} {2002})}\BibitemShut {NoStop}%
\bibitem [{\citenamefont {Shi}\ \emph {et~al.}(2020)\citenamefont {Shi},
  \citenamefont {Zhang},\ and\ \citenamefont {Zhuang}}]{shi2020practical}%
  \BibitemOpen
  \bibfield  {author} {\bibinfo {author} {\bibfnamefont {H.}~\bibnamefont
  {Shi}}, \bibinfo {author} {\bibfnamefont {Z.}~\bibnamefont {Zhang}}, \ and\
  \bibinfo {author} {\bibfnamefont {Q.}~\bibnamefont {Zhuang}},\ }\href@noop {}
  {\bibfield  {journal} {\bibinfo  {journal} {Physical Review Applied}\
  }\textbf {\bibinfo {volume} {13}},\ \bibinfo {pages} {034029} (\bibinfo
  {year} {2020})}\BibitemShut {NoStop}%
\bibitem [{\citenamefont {Zhuang}(2021)}]{zhuang2021quantum}%
  \BibitemOpen
  \bibfield  {author} {\bibinfo {author} {\bibfnamefont {Q.}~\bibnamefont
  {Zhuang}},\ }\href@noop {} {\bibfield  {journal} {\bibinfo  {journal} {arXiv
  preprint arXiv:2103.11054}\ } (\bibinfo {year} {2021})}\BibitemShut {NoStop}%
\bibitem [{\citenamefont {Zhang}\ \emph {et~al.}(2015)\citenamefont {Zhang},
  \citenamefont {Mouradian}, \citenamefont {Wong},\ and\ \citenamefont
  {Shapiro}}]{zhang2015entanglement}%
  \BibitemOpen
  \bibfield  {author} {\bibinfo {author} {\bibfnamefont {Z.}~\bibnamefont
  {Zhang}}, \bibinfo {author} {\bibfnamefont {S.}~\bibnamefont {Mouradian}},
  \bibinfo {author} {\bibfnamefont {F.~N.}\ \bibnamefont {Wong}}, \ and\
  \bibinfo {author} {\bibfnamefont {J.~H.}\ \bibnamefont {Shapiro}},\
  }\href@noop {} {\bibfield  {journal} {\bibinfo  {journal} {Physical review
  letters}\ }\textbf {\bibinfo {volume} {114}},\ \bibinfo {pages} {110506}
  (\bibinfo {year} {2015})}\BibitemShut {NoStop}%
\bibitem [{\citenamefont {Hao}\ \emph {et~al.}(2021)\citenamefont {Hao},
  \citenamefont {Shi}, \citenamefont {Li}, \citenamefont {Shapiro},
  \citenamefont {Zhuang},\ and\ \citenamefont {Zhang}}]{hao2021entanglement}%
  \BibitemOpen
  \bibfield  {author} {\bibinfo {author} {\bibfnamefont {S.}~\bibnamefont
  {Hao}}, \bibinfo {author} {\bibfnamefont {H.}~\bibnamefont {Shi}}, \bibinfo
  {author} {\bibfnamefont {W.}~\bibnamefont {Li}}, \bibinfo {author}
  {\bibfnamefont {J.~H.}\ \bibnamefont {Shapiro}}, \bibinfo {author}
  {\bibfnamefont {Q.}~\bibnamefont {Zhuang}}, \ and\ \bibinfo {author}
  {\bibfnamefont {Z.}~\bibnamefont {Zhang}},\ }\href@noop {} {\bibfield
  {journal} {\bibinfo  {journal} {Physical Review Letters}\ }\textbf {\bibinfo
  {volume} {126}},\ \bibinfo {pages} {250501} (\bibinfo {year}
  {2021})}\BibitemShut {NoStop}%
\bibitem [{\citenamefont {Guha}(2009)}]{guha2009receiver}%
  \BibitemOpen
  \bibfield  {author} {\bibinfo {author} {\bibfnamefont {S.}~\bibnamefont
  {Guha}},\ }in\ \href@noop {} {\emph {\bibinfo {booktitle} {2009 IEEE
  International Symposium on Information Theory}}}\ (\bibinfo {organization}
  {IEEE},\ \bibinfo {year} {2009})\ pp.\ \bibinfo {pages}
  {963--967}\BibitemShut {NoStop}%
\bibitem [{\citenamefont {Djordjevic}(0114)}]{djordjevic2021quantum}%
  \BibitemOpen
  \bibfield  {author} {\bibinfo {author} {\bibfnamefont {I.~B.}\ \bibnamefont
  {Djordjevic}},\ }\href {\doibase 10.1109/JPHOT.2021.3075881} {\bibfield
  {journal} {\bibinfo  {journal} {IEEE Photonics Journal}\ }\textbf {\bibinfo
  {volume} {13}},\ \bibinfo {pages} {1} (\bibinfo {year} {June 2021, Article
  ID: 7500114})}\BibitemShut {NoStop}%
\bibitem [{\citenamefont
  {Djordjevic}(2021{\natexlab{a}})}]{djordjevic2021entanglement}%
  \BibitemOpen
  \bibfield  {author} {\bibinfo {author} {\bibfnamefont {I.~B.}\ \bibnamefont
  {Djordjevic}},\ }\href {10.1109/ACCESS.2021.3137670} {\bibfield  {journal}
  {\bibinfo  {journal} {IEEE Access}\ }\textbf {\bibinfo {volume} {9}},\
  \bibinfo {pages} {168930} (\bibinfo {year} {2021}{\natexlab{a}})}\BibitemShut
  {NoStop}%
\bibitem [{\citenamefont {Bhadani}(2021)}]{bhadani2021optimal}%
  \BibitemOpen
  \bibfield  {author} {\bibinfo {author} {\bibfnamefont {R.}~\bibnamefont
  {Bhadani}},\ }\href@noop {} {\enquote {\bibinfo {title} {Optimal receiver
  design for quantum communication},}\ } (\bibinfo {year} {2021})\BibitemShut
  {NoStop}%
\bibitem [{\citenamefont {Plenio}\ and\ \citenamefont
  {Vedral}(1998)}]{plenio1998teleportation}%
  \BibitemOpen
  \bibfield  {author} {\bibinfo {author} {\bibfnamefont {M.~B.}\ \bibnamefont
  {Plenio}}\ and\ \bibinfo {author} {\bibfnamefont {V.}~\bibnamefont
  {Vedral}},\ }\href@noop {} {\bibfield  {journal} {\bibinfo  {journal}
  {Contemporary physics}\ }\textbf {\bibinfo {volume} {39}},\ \bibinfo {pages}
  {431} (\bibinfo {year} {1998})}\BibitemShut {NoStop}%
\bibitem [{\citenamefont {Shao-Ming}\ \emph {et~al.}(2010)\citenamefont
  {Shao-Ming}, \citenamefont {Albeverio}, \citenamefont {Cabello},
  \citenamefont {Jing},\ and\ \citenamefont {Goswami}}]{shao2010quantum}%
  \BibitemOpen
  \bibfield  {author} {\bibinfo {author} {\bibfnamefont {F.}~\bibnamefont
  {Shao-Ming}}, \bibinfo {author} {\bibfnamefont {S.}~\bibnamefont
  {Albeverio}}, \bibinfo {author} {\bibfnamefont {A.}~\bibnamefont {Cabello}},
  \bibinfo {author} {\bibfnamefont {N.}~\bibnamefont {Jing}}, \ and\ \bibinfo
  {author} {\bibfnamefont {D.}~\bibnamefont {Goswami}},\ }\href@noop {}
  {\bibfield  {journal} {\bibinfo  {journal} {Advances in Mathematical
  Physics}\ }\textbf {\bibinfo {volume} {2010}} (\bibinfo {year}
  {2010})}\BibitemShut {NoStop}%
\bibitem [{\citenamefont {Eisert}(2006)}]{eisert2006entanglement}%
  \BibitemOpen
  \bibfield  {author} {\bibinfo {author} {\bibfnamefont {J.}~\bibnamefont
  {Eisert}},\ }\href@noop {} {\bibfield  {journal} {\bibinfo  {journal} {arXiv
  preprint quant-ph/0610253}\ } (\bibinfo {year} {2006})}\BibitemShut {NoStop}%
\bibitem [{\citenamefont {Miller}(2011)}]{miller2011entanglement}%
  \BibitemOpen
  \bibfield  {author} {\bibinfo {author} {\bibfnamefont {J.}~\bibnamefont
  {Miller}},\ }\href@noop {} {\bibfield  {journal} {\bibinfo  {journal}
  {Physics Today}\ }\textbf {\bibinfo {volume} {64}},\ \bibinfo {pages} {15}
  (\bibinfo {year} {2011})}\BibitemShut {NoStop}%
\bibitem [{\citenamefont {Bennett}\ \emph {et~al.}(1999)\citenamefont
  {Bennett}, \citenamefont {Shor}, \citenamefont {Smolin},\ and\ \citenamefont
  {Thapliyal}}]{bennett1999entanglement}%
  \BibitemOpen
  \bibfield  {author} {\bibinfo {author} {\bibfnamefont {C.~H.}\ \bibnamefont
  {Bennett}}, \bibinfo {author} {\bibfnamefont {P.~W.}\ \bibnamefont {Shor}},
  \bibinfo {author} {\bibfnamefont {J.~A.}\ \bibnamefont {Smolin}}, \ and\
  \bibinfo {author} {\bibfnamefont {A.~V.}\ \bibnamefont {Thapliyal}},\
  }\href@noop {} {\bibfield  {journal} {\bibinfo  {journal} {Physical Review
  Letters}\ }\textbf {\bibinfo {volume} {83}},\ \bibinfo {pages} {3081}
  (\bibinfo {year} {1999})}\BibitemShut {NoStop}%
\bibitem [{\citenamefont {Giovannetti}\ \emph {et~al.}(2003)\citenamefont
  {Giovannetti}, \citenamefont {Lloyd}, \citenamefont {Maccone},\ and\
  \citenamefont {Shor}}]{giovannetti2003entanglement}%
  \BibitemOpen
  \bibfield  {author} {\bibinfo {author} {\bibfnamefont {V.}~\bibnamefont
  {Giovannetti}}, \bibinfo {author} {\bibfnamefont {S.}~\bibnamefont {Lloyd}},
  \bibinfo {author} {\bibfnamefont {L.}~\bibnamefont {Maccone}}, \ and\
  \bibinfo {author} {\bibfnamefont {P.~W.}\ \bibnamefont {Shor}},\ }\href@noop
  {} {\bibfield  {journal} {\bibinfo  {journal} {Physical review letters}\
  }\textbf {\bibinfo {volume} {91}},\ \bibinfo {pages} {047901} (\bibinfo
  {year} {2003})}\BibitemShut {NoStop}%
\bibitem [{\citenamefont {Cubitt}\ \emph {et~al.}(2010)\citenamefont {Cubitt},
  \citenamefont {Leung}, \citenamefont {Matthews},\ and\ \citenamefont
  {Winter}}]{cubitt2010improving}%
  \BibitemOpen
  \bibfield  {author} {\bibinfo {author} {\bibfnamefont {T.~S.}\ \bibnamefont
  {Cubitt}}, \bibinfo {author} {\bibfnamefont {D.}~\bibnamefont {Leung}},
  \bibinfo {author} {\bibfnamefont {W.}~\bibnamefont {Matthews}}, \ and\
  \bibinfo {author} {\bibfnamefont {A.}~\bibnamefont {Winter}},\ }\href@noop {}
  {\bibfield  {journal} {\bibinfo  {journal} {Physical Review Letters}\
  }\textbf {\bibinfo {volume} {104}},\ \bibinfo {pages} {230503} (\bibinfo
  {year} {2010})}\BibitemShut {NoStop}%
\bibitem [{\citenamefont {Lopez-Mago}(2012)}]{lopez2012implementation}%
  \BibitemOpen
  \bibfield  {author} {\bibinfo {author} {\bibfnamefont {D.}~\bibnamefont
  {Lopez-Mago}},\ }\emph {\bibinfo {title} {Implementation of a two-photon
  michelson interferometer for quantum optical coherence tomography}},\
  \href@noop {} {Ph.D. thesis},\ \bibinfo  {school} {Instituto Technologico y
  de Estudios Superiores de Monterrey} (\bibinfo {year} {2012})\BibitemShut
  {NoStop}%
\bibitem [{\citenamefont {Christ}\ \emph {et~al.}(2013)\citenamefont {Christ},
  \citenamefont {Fedrizzi}, \citenamefont {Hübel}, \citenamefont {Jennewein},\
  and\ \citenamefont {Silberhorn}}]{CHRIST2013351}%
  \BibitemOpen
  \bibfield  {author} {\bibinfo {author} {\bibfnamefont {A.}~\bibnamefont
  {Christ}}, \bibinfo {author} {\bibfnamefont {A.}~\bibnamefont {Fedrizzi}},
  \bibinfo {author} {\bibfnamefont {H.}~\bibnamefont {Hübel}}, \bibinfo
  {author} {\bibfnamefont {T.}~\bibnamefont {Jennewein}}, \ and\ \bibinfo
  {author} {\bibfnamefont {C.}~\bibnamefont {Silberhorn}},\ }in\ \href
  {\doibase https://doi.org/10.1016/B978-0-12-387695-9.00011-1} {\emph
  {\bibinfo {booktitle} {Single-Photon Generation and Detection}}},\ \bibinfo
  {series} {Experimental Methods in the Physical Sciences}, Vol.~\bibinfo
  {volume} {45},\ \bibinfo {editor} {edited by\ \bibinfo {editor}
  {\bibfnamefont {A.}~\bibnamefont {Migdall}}, \bibinfo {editor} {\bibfnamefont
  {S.~V.}\ \bibnamefont {Polyakov}}, \bibinfo {editor} {\bibfnamefont
  {J.}~\bibnamefont {Fan}}, \ and\ \bibinfo {editor} {\bibfnamefont {J.~C.}\
  \bibnamefont {Bienfang}}}\ (\bibinfo  {publisher} {Academic Press},\ \bibinfo
  {year} {2013})\ pp.\ \bibinfo {pages} {351--410}\BibitemShut {NoStop}%
\bibitem [{\citenamefont {Mauerer}\ \emph {et~al.}(2009)\citenamefont
  {Mauerer}, \citenamefont {Avenhaus}, \citenamefont {Helwig},\ and\
  \citenamefont {Silberhorn}}]{mauerer2009colors}%
  \BibitemOpen
  \bibfield  {author} {\bibinfo {author} {\bibfnamefont {W.}~\bibnamefont
  {Mauerer}}, \bibinfo {author} {\bibfnamefont {M.}~\bibnamefont {Avenhaus}},
  \bibinfo {author} {\bibfnamefont {W.}~\bibnamefont {Helwig}}, \ and\ \bibinfo
  {author} {\bibfnamefont {C.}~\bibnamefont {Silberhorn}},\ }\href@noop {}
  {\bibfield  {journal} {\bibinfo  {journal} {Physical Review A}\ }\textbf
  {\bibinfo {volume} {80}},\ \bibinfo {pages} {053815} (\bibinfo {year}
  {2009})}\BibitemShut {NoStop}%
\bibitem [{\citenamefont {Wilde}(2013)}]{wilde2013quantum}%
  \BibitemOpen
  \bibfield  {author} {\bibinfo {author} {\bibfnamefont {M.~M.}\ \bibnamefont
  {Wilde}},\ }\href@noop {} {\emph {\bibinfo {title} {Quantum information
  theory}}}\ (\bibinfo  {publisher} {Cambridge University Press},\ \bibinfo
  {year} {2013})\BibitemShut {NoStop}%
\bibitem [{\citenamefont {Ou}(2017)}]{ou2017quantum}%
  \BibitemOpen
  \bibfield  {author} {\bibinfo {author} {\bibfnamefont {Z.~J.}\ \bibnamefont
  {Ou}},\ }\href@noop {} {\emph {\bibinfo {title} {Quantum optics for
  experimentalists}}}\ (\bibinfo  {publisher} {World Scientific Publishing
  Company},\ \bibinfo {year} {2017})\BibitemShut {NoStop}%
\bibitem [{\citenamefont {Lefevre}(1989)}]{lefevre1989power}%
  \BibitemOpen
  \bibfield  {author} {\bibinfo {author} {\bibfnamefont {D.}~\bibnamefont
  {Lefevre}},\ }\href@noop {} {\bibfield  {journal} {\bibinfo  {journal}
  {Cynetics corporation}\ } (\bibinfo {year} {1989})}\BibitemShut {NoStop}%
\bibitem [{\citenamefont {Tan}\ \emph {et~al.}(2008)\citenamefont {Tan},
  \citenamefont {Erkmen}, \citenamefont {Giovannetti}, \citenamefont {Guha},
  \citenamefont {Lloyd}, \citenamefont {Maccone}, \citenamefont {Pirandola},\
  and\ \citenamefont {Shapiro}}]{tan2008quantum}%
  \BibitemOpen
  \bibfield  {author} {\bibinfo {author} {\bibfnamefont {S.-H.}\ \bibnamefont
  {Tan}}, \bibinfo {author} {\bibfnamefont {B.~I.}\ \bibnamefont {Erkmen}},
  \bibinfo {author} {\bibfnamefont {V.}~\bibnamefont {Giovannetti}}, \bibinfo
  {author} {\bibfnamefont {S.}~\bibnamefont {Guha}}, \bibinfo {author}
  {\bibfnamefont {S.}~\bibnamefont {Lloyd}}, \bibinfo {author} {\bibfnamefont
  {L.}~\bibnamefont {Maccone}}, \bibinfo {author} {\bibfnamefont
  {S.}~\bibnamefont {Pirandola}}, \ and\ \bibinfo {author} {\bibfnamefont
  {J.~H.}\ \bibnamefont {Shapiro}},\ }\href@noop {} {\bibfield  {journal}
  {\bibinfo  {journal} {Physical review letters}\ }\textbf {\bibinfo {volume}
  {101}},\ \bibinfo {pages} {253601} (\bibinfo {year} {2008})}\BibitemShut
  {NoStop}%
\bibitem [{\citenamefont {Guan}\ \emph {et~al.}(2017)\citenamefont {Guan},
  \citenamefont {Ma}, \citenamefont {Shi}, \citenamefont {Zhu}, \citenamefont
  {Younce}, \citenamefont {Chen}, \citenamefont {Roman}, \citenamefont {Ophir},
  \citenamefont {Liu}, \citenamefont {Ding} \emph {et~al.}}]{guan2017compact}%
  \BibitemOpen
  \bibfield  {author} {\bibinfo {author} {\bibfnamefont {H.}~\bibnamefont
  {Guan}}, \bibinfo {author} {\bibfnamefont {Y.}~\bibnamefont {Ma}}, \bibinfo
  {author} {\bibfnamefont {R.}~\bibnamefont {Shi}}, \bibinfo {author}
  {\bibfnamefont {X.}~\bibnamefont {Zhu}}, \bibinfo {author} {\bibfnamefont
  {R.}~\bibnamefont {Younce}}, \bibinfo {author} {\bibfnamefont
  {Y.}~\bibnamefont {Chen}}, \bibinfo {author} {\bibfnamefont {J.}~\bibnamefont
  {Roman}}, \bibinfo {author} {\bibfnamefont {N.}~\bibnamefont {Ophir}},
  \bibinfo {author} {\bibfnamefont {Y.}~\bibnamefont {Liu}}, \bibinfo {author}
  {\bibfnamefont {R.}~\bibnamefont {Ding}},  \emph {et~al.},\ }\href@noop {}
  {\bibfield  {journal} {\bibinfo  {journal} {Optics Express}\ }\textbf
  {\bibinfo {volume} {25}},\ \bibinfo {pages} {28957} (\bibinfo {year}
  {2017})}\BibitemShut {NoStop}%
\bibitem [{\citenamefont {Shapiro}\ \emph {et~al.}(2005)\citenamefont
  {Shapiro}, \citenamefont {Guha},\ and\ \citenamefont
  {Erkmen}}]{shapiro2005ultimate}%
  \BibitemOpen
  \bibfield  {author} {\bibinfo {author} {\bibfnamefont {J.}~\bibnamefont
  {Shapiro}}, \bibinfo {author} {\bibfnamefont {S.}~\bibnamefont {Guha}}, \
  and\ \bibinfo {author} {\bibfnamefont {B.}~\bibnamefont {Erkmen}},\
  }\href@noop {} {\bibfield  {journal} {\bibinfo  {journal} {Journal of Optical
  Networking}\ }\textbf {\bibinfo {volume} {4}},\ \bibinfo {pages} {501}
  (\bibinfo {year} {2005})}\BibitemShut {NoStop}%
\bibitem [{\citenamefont {Takeoka}\ and\ \citenamefont
  {Guha}(2014)}]{takeoka2014capacity}%
  \BibitemOpen
  \bibfield  {author} {\bibinfo {author} {\bibfnamefont {M.}~\bibnamefont
  {Takeoka}}\ and\ \bibinfo {author} {\bibfnamefont {S.}~\bibnamefont {Guha}},\
  }\href@noop {} {\bibfield  {journal} {\bibinfo  {journal} {Physical Review
  A}\ }\textbf {\bibinfo {volume} {89}},\ \bibinfo {pages} {042309} (\bibinfo
  {year} {2014})}\BibitemShut {NoStop}%
\bibitem [{\citenamefont
  {Djordjevic}(2021{\natexlab{b}})}]{djordjevic2021entanglement2}%
  \BibitemOpen
  \bibfield  {author} {\bibinfo {author} {\bibfnamefont {I.~B.}\ \bibnamefont
  {Djordjevic}},\ }\href@noop {} {\bibfield  {journal} {\bibinfo  {journal}
  {IEEE Access}\ }\textbf {\bibinfo {volume} {9}},\ \bibinfo {pages} {42604}
  (\bibinfo {year} {2021}{\natexlab{b}})}\BibitemShut {NoStop}%
\bibitem [{\citenamefont {Shieh}\ and\ \citenamefont
  {Djordjevic}(2009)}]{SHIEH2009263}%
  \BibitemOpen
  \bibfield  {author} {\bibinfo {author} {\bibfnamefont {W.}~\bibnamefont
  {Shieh}}\ and\ \bibinfo {author} {\bibfnamefont {I.}~\bibnamefont
  {Djordjevic}},\ }in\ \href {\doibase
  https://doi.org/10.1016/B978-0-12-374879-9.00007-1} {\emph {\bibinfo
  {booktitle} {OFDM for Optical Communications}}},\ \bibinfo {editor} {edited
  by\ \bibinfo {editor} {\bibfnamefont {W.}~\bibnamefont {Shieh}}\ and\
  \bibinfo {editor} {\bibfnamefont {I.}~\bibnamefont {Djordjevic}}}\ (\bibinfo
  {publisher} {Academic Press},\ \bibinfo {address} {Oxford},\ \bibinfo {year}
  {2009})\ pp.\ \bibinfo {pages} {263--294}\BibitemShut {NoStop}%
\end{thebibliography}%
\end{document}